\documentclass[lettersize,journal]{IEEEtran}
\usepackage{amsmath,amsfonts}
\usepackage{algorithmic}
\usepackage{algorithm}
\usepackage{array}
\usepackage{textcomp}
\usepackage{stfloats}
\usepackage{url}
\usepackage{verbatim}
\usepackage{graphicx}
\usepackage{cite}
\usepackage{soul}
\usepackage{blindtext}
\usepackage{placeins}
\usepackage[utf8]{inputenc}
\usepackage{bm}
\usepackage{multirow}
\usepackage{subfigure}
\usepackage{tabularx,lipsum,booktabs}
\usepackage{soulutf8}
\usepackage{caption}
\usepackage{listings}
\usepackage{amssymb}
\usepackage[breaklinks=true]{hyperref}
\usepackage{multicol}
\usepackage{comment}
\usepackage[version=4]{mhchem}
\newsavebox{\bigimage}
\usepackage[dvipsnames,svgnames]{xcolor}

\usepackage{siunitx}

\begin{document}

\title{Internet of Paint (IoP): Channel Modeling and Capacity Analysis for Terahertz Electromagnetic Nanonetworks Embedded in Paint}
\author{Lasantha~Thakshila~Wedage,~\IEEEmembership{Student Member,~IEEE}, Mehmet C. Vuran,~\IEEEmembership{Member,~IEEE}, Bernard~Butler\thanks{\it{Lasantha Thakshila Wedage and Bernard Butler are with the Walton Institute, South East Technological University, Ireland.}},\thanks{\it{Mehmet C. Vuran and Sasitharan Balasubramaniam are with the University of Nebraska-Lincoln, USA.}} Yevgeni~Koucheryavy,~\IEEEmembership{Senior Member,~IEEE} \thanks{\it{Yevgeni Koucheryavy is with the Tampere University of Technology, Finland.}} and Sasitharan~Balasubramaniam,~\IEEEmembership{Senior Member,~IEEE}}

\maketitle

\begin{abstract}
This work opens a new chapter in the $100,000$ year-old concept of paint, by leveraging innovations in nano-technology in the sub-THz frequency range. More specifically, the groundbreaking concept of Internet of Paint (IoP) is introduced along with a comprehensive channel model and a capacity analysis for nano-scale radios embedded in paint and communicating through paint. Nano-network devices, integrated within a paint medium, communicate via a multipath strategy, encompassing direct waves, reflections from interfaces, and lateral wave propagation. The evaluation incorporates three distinct paint types to assess path losses, received powers, and channel capacity. Analysis of path loss indicates a slight non-linear increase with both frequency and Line of Sight (LoS) distance between transceivers. Notably, paints with high refractive indexes result in the highest path loss. Moreover, burying transceivers at similar depths near the Air-Paint interface showcases promising performance of lateral waves with increasing LoS distance. Increasing paint layer depth leads to amplified attenuation, while total received power exhibits promising results when in close proximity to the Air-Paint interface but steeply declines with burial depth. Additionally, a substantial reduction in channel capacity is observed with LoS distance and burial depth, so transceivers need to be close together and in proximity of the A-P interface to communicate effectively. Comparing paint and air mediums, IoP demonstrates approximately two orders of magnitude reduction in channel capacity compared to air-based communication channels. This paper provides valuable insights into the potential of IoP communication within paint mediums and offers a foundation for further advancements in this emerging field.
\end{abstract}

\begin{IEEEkeywords}
THz communication, Internet of Paint (IoP), Nanonetworks, Channel model, Channel Capacity.
\end{IEEEkeywords}

\section{Introduction}
\IEEEPARstart{N}{ano}-network device capabilities are developing rapidly, due to advances in nanomaterials and components that can be integrated into miniature devices that will communicate in the Terahertz (THz) ($0.1-10$\,THz) frequency spectrum \cite{Nano_network_2021}. Currently, researchers are proposing the use of graphene-based antennas in nano-network devices for limited close-form applications to study Electro-Magnetic (EM) communication characteristics such as on-chip \cite{llatser2013graphene} and intra-body \cite{elayan2017terahertz,vizziello2023intra} communication. The ability to deploy a massive number of nano-scale devices relatively cheaply, within limited areas promises novel applications of \textit{engineering the wireless channel} through Reconfigurable Intelligent Surfaces (RIS) \cite{jornet2023nanonetworking, Wedage_2023} and \textit{RF-based sensing} for non-intrusive low-cost sensing applications such as activity detection, movement detection, and ambient sensing \cite{THz_detectors, kianoush2019passive, dai2009remote}.

In the near future, nano-scale transceivers are expected to be low-power, small-size, and low-cost, opening up the possibility of such devices becoming pervasive in indoor environments. For example, the RIS that can be embedded into indoor environments could be used to reflect signals to compensate for channel impairments due to blockages (e.g., furniture, people) in the environment. Moreover, there have been proposals to place such devices on the surface of walls mainly for sensing purposes \cite{sensing2019room}, and can be extended to enhance reliable communication channels in the future. However, existing solutions are intrusive and visually unappealing, adversely affecting both the functional properties and the appearance of indoor walls. In this paper, we introduce the novel concept of \textbf{\textit{Internet of Paint (IoP)}}, where nano-devices are integrated into the paint mixture and applied to the wall, enabling numerous applications (e.g., sensing or wall-integrated RIS), while also providing visually appealing and practical integration of nano-networks to make such deployments \textit{seamless}. To this end, we mainly focus on analyzing the nano-device communication through the paint medium utilizing THz links. One of the possible application areas of IoP is IRS/RIS applications. More specifically, the developed channel model could be used to support intra-signaling between elements of IRS/RIS devices that can also reflect from the wall.

\begin{figure*}[ht]
    \centering
    \includegraphics[width=0.85\linewidth]{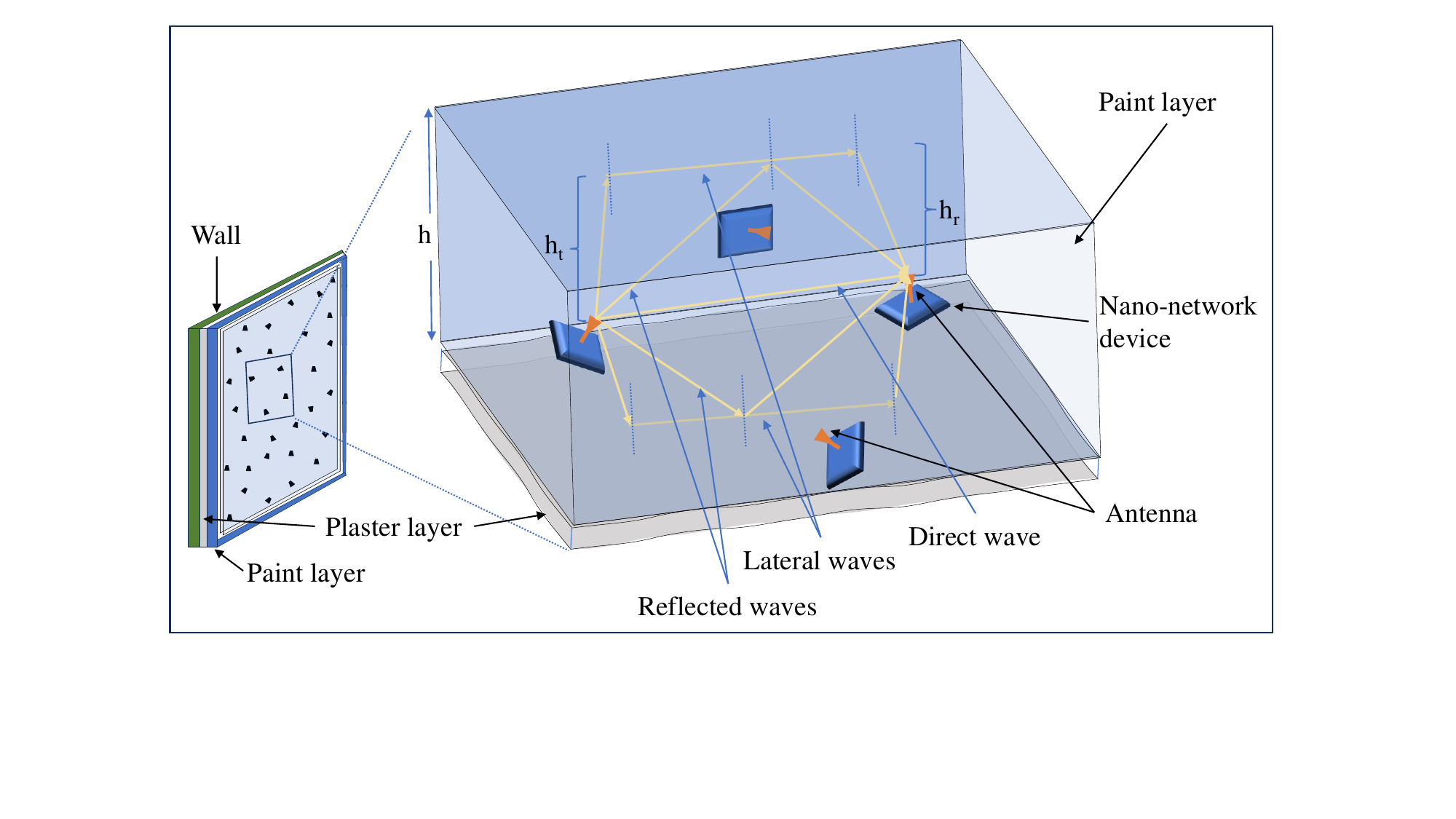}
    \caption{Internet of Paint (IoP) where nano-devices are embedded into the paint mixture and applied onto a wall, enabling nano-network communication at short range using THz frequencies.}
    \label{fig:3D_overall_model}
\end{figure*}

Paint has been an integral part of human life dating back $100,000$ years \cite{pappa2011oldest,sample2011stone}. Conventionally, paint consists of color pigments suspended in a liquid mixture. When paint is applied to a solid surface, it adds a film-like layer that can be used to protect a large surface like a wall, in order to enhance the interior or exterior architecture, or to create an artwork such as a painting. More recently, pigment-free \textit{structural} paints have been developed with the aid of nanoparticle technology. For example, in \cite{cencillo2023ultralight}, sub-wavelength plasmonic cavities mixed in paint composition can result in significant reduction of weight compared to conventional paint with micro-scale pigments. In this paper, we envision a next-generation paint structure that is infused with nano-scale communication elements. Such a paint mixture will significantly ease installation of RIS or ambient sensing solutions without impacting the aesthetics of interior environments. More specifically, rather than affixing miniature transceivers to an indoor surface, we envision that the nano-devices are embedded in the paint mixture before it is applied to the wall. This has the dual benefit of supporting and protecting the resulting nano-network by the paint layer. 
This integrated nano-network within the paint layer on the wall (see Fig. \ref{fig:3D_overall_model}) leads to the novel concept of IoP, which is the focus of this paper.

While the concept of IoP nano-networks is intriguing, it also introduces several challenges. The placement and orientation of the devices are not controlled in the paint layer. Relative to a fixed nano-device network, our proposal would introduce more devices to ensure a relatively dense network with fewer holes in the network coverage. Also, the paint layer attenuates the received signal strength, reducing the effective range and/or the channel capacity of each link. Thermal noise generated by the nano-devices, with its possible effect on communication, is not considered in this study. In this paper, we propose an IoP channel model (see Section~\ref{sec:channel_model}) that captures the five canonical signal paths between devices suspended in a paint layer. Section~\ref{sec:channel_capacity} presents a channel capacity model for multi-path communication for a typical link. Extensive numerical evaluations are performed to validate the channel model---these are reported in Section~\ref{sec:Results_and_Discussion}. In particular, numerical experiments explore received signal strength and path loss relative to the transmitted signal strength, and  (Section~\ref{subsec:PL_and_RP}), considers the effects of different transceiver burial depths (Section~\ref{subsec:Burial_Depth_Analysis}), and how channel capacity is affected by transmission distance and paint mixture types (Section~\ref{subsec:Channel_Capacity_Analysis}). Finally, Section~\ref{sec:Conclusion} summarises the contributions and concludes the paper.

\section{Background}
The latest advancements in technology have opened vast opportunities to establish EM communication between nano-network devices in the THz band, enabling them to infer the communication characteristics of various mediums of the EM waves. For instance, many recent studies have investigated the THz EM wave propagation characteristics through human blood \cite{gomez2023optimizing}, tissue \cite{reddy2023photothermal}, and the body \cite{elayan2017terahertz}. Another emerging field is on-chip nano-network communication in various environments using graphene-based plasmonic antennas in the millimeter wave and THz bands \cite{llatser2013graphene, timoneda2018millimeter, abadal2019opportunistic, abadal2019wave}.

Furthermore, there are many studies that investigate wireless communication through harsh environments, such as soil (underground communication) \cite{vuran2018internet, dworak2017terahertz}, water (underwater communication) \cite{kaushal2016underwater}, and dusty mediums \cite{Wedage_comparative_2023}. Signal propagating behavior significantly varies with the medium properties, particularly in the THz spectrum. As a result, each environment necessitates the development of specific channel models to predict the propagation characteristics of THz EM waves. To mitigate signal strength attenuation, directional antennas are recommended for communication in the THz spectrum \cite{hossain2019stochastic}, with beam steering \cite{monnai2023terahertz} as needed. Since we cannot control the orientation of the nano-devices when they have been mixed into the paint, omni-directional antennas are the most appropriate choice. Furthermore, the short communication distances between the nano-devices in a dense nanonetwork favors the existence of multi-path communication, so the channel model needs to account for this.  

Paint composition includes pigments, binders, solvents (liquids), and additives. The various types of pigments provide color for the paint, while binders work to ``bind" the mixture and create the paint film. Communication through paint utilizing THz-enabled embedded nanonetwork devices is a new frontier for wireless communications. As far as the authors are aware, this is the first study that discusses channel modeling and capacity analysis for IoP, utilizing THz spectrum to establish communication between nano-devices. This exploration promises insights into the dynamics of nano-network communication via the proposed \emph{Internet of Paint (IoP)}, paving the way for establishing novel communication channels in the THz spectrum with significant channel capacities while offering scope for a novel realization of smart walls.

\begin{figure*}[ht]
    \centering
    \includegraphics[width=0.85\linewidth]{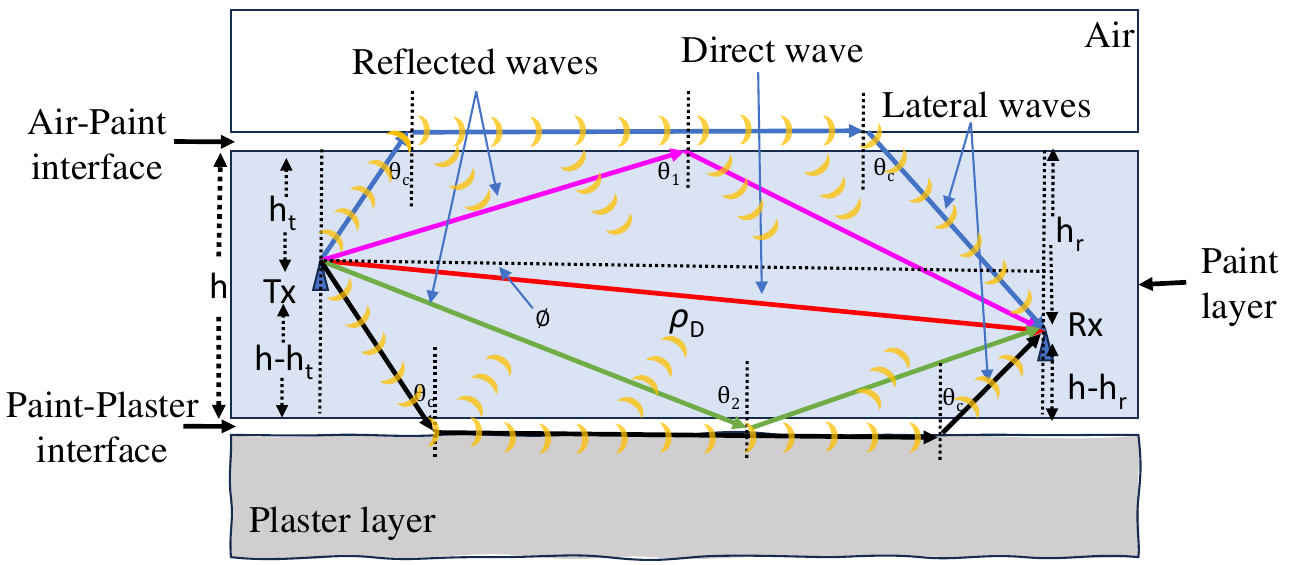}
    \caption{Illustration of nano-devices embedded in paint resulting in direct path, reflected paths, as well as lateral waves that can propagate along the air-path as well as paint-plaster interface. 
    }
    \label{fig:Channel_model}
\end{figure*}

\section{Internet of paint (IoP) Channel Model}
\label{sec:channel_model}
In the following, we present a channel model for communication between nano-devices embedded in paint. We envision that these IoP devices will be constructed from nanomaterials, and this includes components that can harvest energy. An example is the use of nano components that deform mechanically to harvest energy (e.g., compression of zinc oxide nanowires \cite{wang2008towards, xu2010piezoelectric}). The compression can be achieved using vibration within the paint layer caused by an external ultrasound source, which has been proven to vibrate piezoelectric crystals in other media \cite{seo2016wireless}. It might also be possible to reduce the energy demand per transceiver. If the nanotransceivers cooperate by sharing their communication capacity, it could be aggregated over large numbers of transceivers to meet bursty communication demands, and/or the average duty cycle per nanotransceiver could be reduced when data rate requirements are lower. In future work, we will develop our ideas regarding how the IoP nanonetwork would be powered in a sustainable fashion.

Moreover, nano-devices should be sealed properly to avoid chemical reactions with the environment before mixing with paint. This is especially important for devices made using graphene. For example, a Teflon substrate can be used in combination with graphene to provide additional protection against chemical reactions \cite{khan2020high}. After applying, paint can act as a protective layer for nano-devices, safeguarding them from moisture and other environmental factors. 

In the THz band, channel behavior is often tightly coupled with antenna behavior because highly directional antennas are needed to increase the communication range. This is mainly due to the high attenuation in this frequency band. The directionality of the antennas can be achieved through electronically steerable antenna arrays. However, the design of antenna arrays embedded in mediums other than air is challenging because of the unique interplay between antenna impedance, wavelength, and effects from interfaces~\cite{salam2017smart}. More importantly, the design of such arrays for the THz spectrum in a paint medium is an open issue. Furthermore, the design of highly directive antennas for IoP requires in-depth knowledge of the channel characteristics. Therefore, we limit the scope of this paper to the analysis of the IoP channel. Accordingly, an idealized point source within the paint medium is considered. As we discuss in Section~\ref{sec:Results_and_Discussion}, this approach leads to unique insights into the design requirements for IoP antennas, which is out of the scope of this paper. 

The IoP channel model is illustrated in Fig.~\ref{fig:Channel_model}. We consider a real-world application where a (relatively) thick layer of paint is applied on a thin plaster surface. Depending on construction approaches in different parts of the world (e.g., Europe vs US), the plaster layer could be replaced by a drywall. This results in three different media for EM wave propagation. These three media are: air, the paint layer, and the plaster layer. By embedding transceivers within the paint layer, the EM waves can propagate in these media with different speeds, due to the differences in the dielectric constants of each medium. Furthermore, the three different layers create two interfaces: Air-Paint (A-P) and Paint-Plaster (P-P). An EM wave that is incident to each of these interfaces undergoes reflection and refraction, which also results in a grouped wave phenomenon called \textit{lateral waves}~\cite{vuran2018internet}.

The approximate thickness of the paint layer ($h$) is assumed to be constant relative to the very short communication range of nano-network devices, $\rho_D$ is the line of sight (LoS)\footnote{We use the conventional term LoS vaguely here to refer to the direct distance because the word `sight` looses its meaning within the visually opaque paint medium.} distance between the transmitter and the receiver in the paint layer, and the burial depths of the transmitter and receiver are $h_t$ and $h_r$ with respect to the Air-Paint interface, respectively. The burial depths of the transceivers are assumed to vary in the interval $(0,h)$, which implies that transceivers will always be inside the paint. The angle between the LoS and the horizontal Air-Paint interface is $\phi$, and it can be determined using $\phi = \arcsin(|h_t-h_r|/\rho_D)$. 

For each medium, we assume its dielectric properties are both homogeneous. We also model three main types of paths, and there are five dominant waves that propagate between the transmitter and the receiver. The three types of paths are: (1) Direct, (2) Reflected, and (3) Lateral paths. Accordingly, the five dominant waves are: (i) direct wave (DW), (ii) reflected wave from the Air-Paint interface (RW-A), (iii) reflected wave from the Paint-Plaster interface (RW-P), (iv) lateral wave through the A-P interface (LW-A), and (v) lateral wave through the P-P interface (LW-P) (see Fig.~\ref{fig:Channel_model}). Note that in this analysis, we ignore effects due to successive reflections between the A-P and P-P interfaces, and the resulting contributions to the two lateral waves, because of the high attenuation in the THz spectrum.

Note that in this analysis, we assume that the receiver can receive incoming waves from every direction. In a setting with directional antennas, this assumption is not valid and one can consider directionality patterns that could amplify desired waves and attenuate others, or beam selection solutions that could select the best path. Accordingly, the results of this analysis paves the way for such algorithmic methods. 

In the following, we illustrate each type of path and derive the received power for each dominant wave. It is important to note that, depending on the differences in the propagation speed within each media, the deployment depths and distances, as well as the capabilities of the radios, each wave could be resolved, or could constructively or destructively combine at the receiver. We initially analyze each wave separately and then evaluate cases when they are combined in Section~\ref{sec:Results_and_Discussion}. 

\subsection{Direct Path}
The path loss corresponding to the LoS communication through the paint can be expressed as the sum of the spreading and absorption losses due to EM wave propagation in the paint medium following~\cite{jornet2011channel,Akyildiz2012terahertz} as: 

\begin{equation}
    PL_D = 20 \log_{10}{\left(\frac{4\pi f \rho_D }{c_p}\right)} + 10\log_{10}{e^{\rho_D K_p(f)}},
    \label{eq:Pathloss_direct}
\end{equation}
where $\rho_D$ is the transmission distance through the paint. The frequency-dependent absorption coefficient of paint ($K_p(f)$) in $100$\,GHz to $500$\,GHz range can be expressed according to the data given in \cite{refractive2008} as,  
\begin{equation}
    \label{eq:kf-paint}
    K_p(f) [cm^{-1}] = 0.0091f + 1.3921.  
\end{equation}
The EM wave speed in the paint medium, $c_p$, is significantly lower than that in the air, $c$, because $c_p \equiv c/n_p$ \cite{salam2020statistical}, where $n_p$ is the refractive index of paint (relative to vacuum).

Therefore, the received power from the DW is given as,

\begin{equation}
    P_r^D (dBm) = P_t (dBm) + G_t (dBi) + G_r (dBi) - PL_D (dBm),
\end{equation}
where $P_t$ is the transmitted power, $G_t$ and $G_r$ are the transmitter and the receiver antenna gains.

\subsection{Reflected Paths}

Two reflected paths are possible in IoP communication due to the A-P and P-P interfaces. The angle of the reflected wave is determined by Snell's Law for the waves with an incident angle greater than the critical angle (i.e. $\theta_c=\sin^{-1}(n_{o}/n_{i})$), where $n_i$ is the refractive index of the incident medium (i.e., paint) and $n_o$ is the refractive index of either the air or the plaster layer. Note that in practice, the refractive index of paint is greater than that of both the air and the plaster layer, resulting in these reflection patterns. 

\subsubsection{Reflected Wave from the Air-Paint Interface}

\begin{figure}[t]
    \centering
    \includegraphics[width=\linewidth]{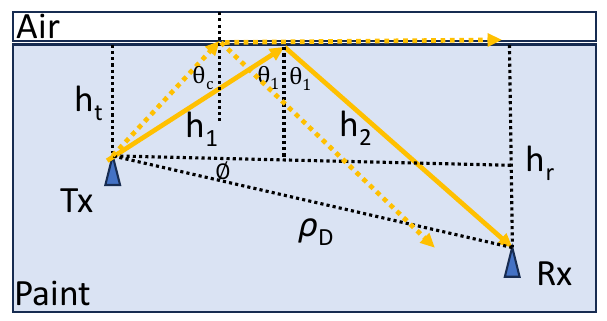}
    \caption{Geometry wave reflected from the Air-Paint interface, indicating specific angles of reflection will result in perfect wave alignment to the receiver.}
    \label{fig:Ref_Air_Paint}
\end{figure}

The geometry of the reflected wave from the Air-Paint (A-P) surface is illustrated in Fig. \ref{fig:Ref_Air_Paint}, where $\phi$ is the angle between LoS and the horizontal A-P interface, and $h_1$ and $h_2$ are the path distances from the reflection point at the A-P interface to the transmitter and the receiver, respectively. The incident angle, $\theta_1$, is determined according to Snell's Law for total internal reflection. 
The path loss corresponding to the reflected wave from the A-P interface, $PL_{R,AP}$, can be expressed following \cite{jornet2011channel}, and the frequency-dependent transfer function of the reflected ray propagation obtained by \cite{han2014multi} as:
\begin{equation}
\label{eqn:reflectedPL}
\begin{split}
    PL_{R,AP}  = & 20 \log_{10}\left(\frac{4\pi f (h_1+h_2) }{c_p}\right) \\
    &+ 10\log_{10}{e^{(h_1+h_2)K_p(f)}} - 10\log_{10}{R_p(f)}\;. 
\end{split}
\end{equation}
 The path lengths can be expressed as $h_1=h_t/\cos(\theta_1)$ and $h_2 = h_r/\cos(\theta_1)$, considering the geometry of the reflected wave according to the Snell's Law, and $R_p(f)$ is the frequency-dependent reflection coefficient considering paint as a rough surface. 

According to the Kirchhoff scattering theory, the reflection coefficient for a rough surface can be obtained by multiplying the smooth surface reflection coefficient, $\gamma_p$, derived from the Fresnel equation with the Rayleigh roughness factor, $\varrho_p(f)$~\cite{tsujimura2017causal} as,
\begin{equation}
    R_p(f) = \gamma_p\cdot\varrho_p(f).
\end{equation}
The Fresnel reflection coefficient ($\gamma_p$) for transverse electric (TE) polarized waves on a smooth paint surface is obtained as:
\begin{equation}
    \gamma_{p} = \dfrac{\cos(\theta_1) - n_p \sqrt{1- \left( \frac{1}{n_p}\sin(\theta_1)\right)^2}}{\cos(\theta_1) + n_p \sqrt{1- \left( \frac{1}{n_p}\sin(\theta_1)\right)^2}},
\end{equation}
where $n_p$ is the refractive index of paint and $\theta_1$ is the angle of incidence~\cite{tsujimura2017causal}.

Moreover, the rough surface height standard deviation ($\sigma_p$ for paint) is a statistical parameter for roughness, which is commonly considered to be Gaussian-distributed. This roughness effect is characterized by a Rayleigh factor \cite{piesiewicz2007scattering} represented as,
\begin{equation}
    \varrho_p(f) = exp\left( - \dfrac{8 \pi^2 f^2 \sigma_p^2 cos^2(\theta_1)}{c_p^2} \right)\;.
\end{equation}

Therefore, the received power of the reflected wave from the Air-Paint interface is given as,
\begin{equation}
    \begin{split}
    P_r^{R,AP} (dBm) =& P_t (dBm) + G_t (dBi) + G_r (dBi) \\
    & - PL_{R,AP} (dBm)\;,    
    \end{split}
\end{equation}
where $P_t$ is the transmitted power, $G_t$ and $G_r$ are the transmitter and the receiver gains.
\begin{figure}[t]
    \centering
    \includegraphics[width=\linewidth]{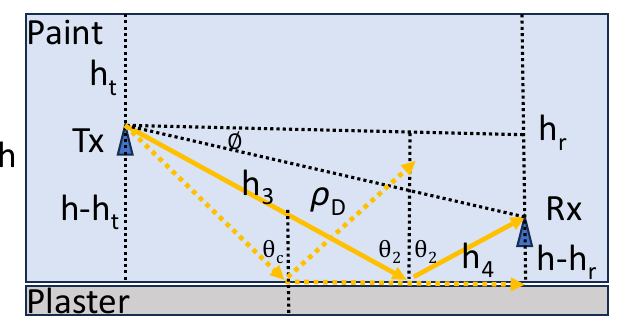}
    \caption{Geometry for the waves reflected from the Paint-Plaster interface, resulting in specific waves that reflect and align to the receiver as well as lateral waves that travel along the paint-plaster interface.}
    \label{fig:Ref_Paint_Plaster}
\end{figure}

\subsubsection{Reflected Wave from the Paint-Plaster Interface}

The geometry of the reflected wave from the Paint-Plaster surface is illustrated in Fig. \ref{fig:Ref_Paint_Plaster}. The incident angle, $\theta_2$, is determined according to the Snell's Law for total internal reflection from the P-P interface.   

The path loss corresponding to the reflected wave from the P-P interface, $PL_{R,PP}$, can be expressed similarly to (\ref{eqn:reflectedPL}), where $h_1$ and $h_2$ are replaced by $h_3=(h-h_t)/\cos(\theta_2)$ and $h_4 = (h-h_r)/\cos(\theta_2)$, respectively. Here, $R_{pl}(f)$ is the frequency-dependent reflection coefficient considering Plaster as a rough surface, which can be found similarly to the explanation in the above subsection by substituting refractive index ($n_{pl}$), rough surface standard deviation ($\sigma_{pl}$), and incident angle ($\theta_2$) for plaster. 

Accordingly, the received power of the reflected wave from the P-P interface is given as,
\begin{equation}
\begin{split}    
    P_r^{R,PP} (dBm) = &P_t (dBm) + G_t (dBi) + G_r (dBi) \\
    &- PL_{R,PP} (dBm)\;.
    \end{split}
\end{equation}
\subsection{Lateral Waves}
\begin{figure}[t]
    \centering
    \includegraphics[width=\linewidth]{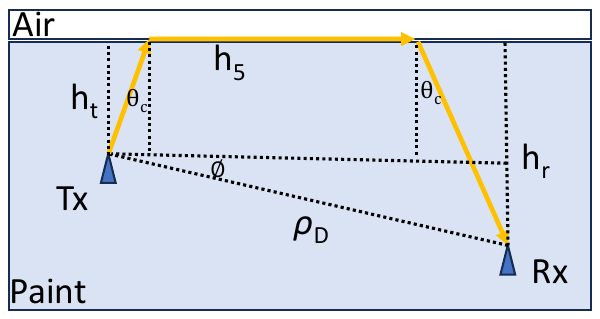}
    \caption{Geometry for refracted waves that results in Lateral waves propagating along the Air-Paint interface, where energy leakage back into the paint resulting in a wave refraction to the receiver.}
    \label{fig:Late_Air_Paint}
\end{figure}

The third type of wave for IoP communication is lateral waves. Such waves occur when EM waves are refracted at the critical angle according to Snell's law~\cite{Clough76}. This phenomenon has been observed and utilized for scenarios involving different media such as underwater~\cite{Wu82}, underground~\cite{vuran2018internet}, ice~\cite{Clough76}, and glass~\cite{Ginzel08}. When a signal source is located inside a high-density medium, close to an interface with a low-density medium (e.g., the earth's surface), the lateral waves originate in the high-density medium (e.g., earth) and propagate through this high-density medium towards the interface crossing over to the medium with lower density (e.g., air), propagate over the lower density medium across the interface, and are received at the higher density medium through refraction (see Fig.~\ref{fig:Late_Air_Paint}). Lateral waves are also called \textit{head waves} as they are generally received first at the receiver because of the higher propagation speed in the lower density medium. Lateral waves occur across a wide frequency spectrum ranging from a few MHz to the optical region. However, we are not aware of a published analysis of lateral waves in the THz band, which follows in Section~\ref{subsec:LW_A_P}. Due to the A-P and P-P interfaces in IoP, two types of lateral waves are considered.

\subsubsection{Lateral Wave along the Air-Paint Interface}
\label{subsec:LW_A_P}

As shown in Fig.~\ref{fig:Late_Air_Paint}, the lateral wave along the air-paint interface (LW-A) originates in the paint medium propagating towards the interface, refracting and propagating through air along the A-P interface, and reaching the receiver in the paint medium. 
 
An accurate characterization of the lateral wave would require experimental evaluations or ray tracing in environments where experimentation is possible~\cite{salam2020statistical}. However, under certain conditions, the lateral wave model can be well approximated by a first-order model, which was validated through experiments in several environments (e.g., forest~\cite{Dence69}). This approximation holds if each medium can be modeled as a half-space. The half-space model certainly holds true for air considering the relative thickness of paint with respect to any room. Furthermore, both the paint and the plaster layers can be approximated by a half-space model considering the relatively short distances considered for nano-scale transceivers and the high attenuation of THz signals.

With this first-order approximation, the lateral wave can be further specified. More specifically, the wave originates at the transmitter in the paint medium and is incident upon the A-P interface at the critical angle of total reflection, $\theta_c$. Then, the wave is refracted into the air and propagates along the A-P interface (over air), leaking energy back into the paint medium at the critical angle of reflection~\cite{Dence69}. As a result, the receiver receives the wave at the critical angle of $\theta_c$. Therefore, the path loss can be expressed as,
\begin{equation} \label{eq:lateral-approx}
\begin{split}
PL_{L,AP}  = &20 \log_{10}{\left[\frac{4\pi f}{c_p} \left(\frac{h_t+h_r}{\cos(\theta_c)} \right)\right]}\\
&+ 10\log_{10}{e^{K_p(f)\left(\frac{h_t+h_r}{\cos(\theta_c)} \right)}}  \\
 &+ 20 \log_{10}{\left(\frac{4\pi f h_5}{c}\right)} + 10\log_{10}{e^{h_5K_{a}(f)}},
\end{split}
\end{equation}

where the first and second terms represent the spreading and absorption losses along the paint paths, and the last two terms capture path loss over air. In (\ref{eq:lateral-approx}), $h_5 = [\rho_D cos(\phi)- (h_t+h_r) tan(\theta_c)]$ is the lateral path distance over air.

The overall molecular absorption coefficient ($K_{a}(f)$) of air at frequency $f$ can be expressed as,
\begin{equation}
    \label{eq:OverallAbs}
    K_{a}(f)=\bold{Q_g} \cdot \bold{K^a_g(f)} ,
\end{equation}

where $\bold{K^a_g(f)}=[K_{g_1}(f), K_{g_2}(f), ..., K_{g_{10}}(f)]$ is the molecular absorption coefficient vector of water vapor ($1\%$) and nine other gases, except for Argon (data corresponding to inert gases is unavailable on the HITRAN database), which can be found in high concentrations in the atmosphere (see Table \ref{tab:AtmosphereComparison}). For the considering gas mixture, the total number of molecules per volume unit (molecules/cm$^3$) vector, $\bold{Q_g}$, at pressure $p$ and temperature $T$, is obtained from the Ideal Gas Law as,

\begin{equation}
    \label{eq:total_number_of_molecules}
    \bold{Q_g}= \frac{p N_a}{R T} \cdot \bold{C_g},
\end{equation}

where $\bold{C_g}=[C_{g_1}(f), C_{g_2}(f), ..., C_{g_{10}}(f)]$ is the gas mixing ratio vector, $N_a$ is the Avogadro constant, and $R$ is the gas constant.

Moreover, $K_g(f)=\sum_{i,g}^{}{k^{i}_{g}(f)}$ and $k^{i}_{g}(f)$ is the monochromatic absorption coefficient of the $i^{th}$ isotopologue of $g^{th}$ gas at frequency $f$. The monochromatic absorption coefficient for each isotopologue of a particular gas in the atmosphere at frequency $f$ is provided in \cite{Hitran2016},
\begin{equation}
    k^{i}_{g}(f) = S^{i}_{g}(T)F^{i}(f),
\end{equation}
where $S^{i}_{g}(T)$ is the line intensity at temperature $T$ referenced to the temperature $296$\,K of the $i^{th}$ isotopologue of $g^{th}$ gas, which can be easily calculated using high-resolution transmission (HITRAN) molecular spectroscopic data. Here, $F^{i}(f)$ is the spectral line shape function at frequency $f$. 
In the lower atmosphere on Earth, including indoor locations, pressure broadening of spectral lines dominates the line shape, and a Lorentz profile can be assumed as the line shape function \cite{Wedage_pathloss_2022,Wedage_comparative_2023}, and it is given by \cite{Hitran2016},
\begin{equation}
\label{eq:F_L}
        F^{i}_L(f)=\dfrac{1}{\pi}\dfrac{\gamma(p,T)}{\gamma(p,T)^2 + [ f - (f^{i}_{g} + \delta(P_{ref}) P)]^2},
    \end{equation}
where $f^{i}_{g}$ is the resonant frequency for isotopologue $i$ of gas $g$, $\gamma(P,T)$ is the Lorentzian (pressure-broadened) Half Width at Half Maximum (HWHM) for gas at pressure $P$ (atm), temperature $T$ (K), and $\delta(P_{ref})$ is the pressure shift at reference pressure ($P_{ref}=1\,atm$). 

 \begin{table}[t]
        \centering
       
        \begin{tabular}{|l|l|}
        \hline
            Gas & Composition  \\
        \hline
          \ce{N2}   & 78.084\%       \\
          \ce{O2}   &  20.946\%     \\
          \ce{Ar}   &  0.93\%    \\
          \ce{H2O}  &  1-3\%      \\
          \ce{CO2}  &  0.003\%   \\
          \ce{CH4}  &  1.5~$ppm$ \\
          \ce{SO2} &  1~$ppm$    \\ 
          \ce{O3}   &  0.05~$ppm$  \\
          \ce{N2O}  & 0.02~$ppm$   \\
          \ce{CO}  &  0.01~$ppm$  \\
          \ce{NH3}  & 0.01~$ppm$ \\
        \hline
        \end{tabular}
         \caption{Atmospheric gas composition on Earth \cite{ep_value} ($ppm$ stands for parts per million).}
        \label{tab:AtmosphereComparison}
    \end{table}

Therefore, the received power from the lateral waves along the A-P interface is given as,
\begin{equation}
\begin{split}    
    P_r^{L,AP} (dBm) =& P_t (dBm) + G_t (dBi) + G_r (dBi)\\
    &- PL_{L,AP} (dB),
\end{split}
\end{equation}
where $P_t$ is the transmitted power, $G_t$ and $G_r$ are the transmitter and the receiver gains.
   
\subsubsection{Lateral Wave along the Paint-Plaster interface}

Since the permittivity of paint is higher than that of the plaster \cite{refractive2008}, the lateral waves also exist through the P-P interface for communication. Therefore, the lateral wave will propagate through both paint and plaster mediums (see Fig. \ref{fig:Ref_Paint_Plaster}). The path loss, $PL_{L,PP}$, can be expressed similar to (\ref{eq:lateral-approx}) by replacing the paint path distance in the first two terms with $(2h-h_t-h_r)/cost(\theta_c)$, $K_a(f)$ with $K_pl(f)$, and $h_6=\rho_D\cos(\phi)-(2h-h_t-h_r)\tan(\theta_c)$. 
$K_{pl}(f)$ is the absorption coefficient of plaster at frequency $f$, which can be expressed following the data given in \cite{refractive2008} for the frequency range of $100$ to $500$\,GHz as,
\begin{equation}
    \label{eq:kf-plaster}
    K_{pl}(f) [cm^{-1}] = 0.00006f^2 - 0.0056f + 0.8819.  
\end{equation}

The EM wave speed in the plaster medium ($c_{pl}$) is $c_{pl} = c/n_{pl}$, where $n_{pl}$ is the refractive index of plaster. 
 
\begin{figure}[t]
    \centering
    \includegraphics[width=\linewidth]{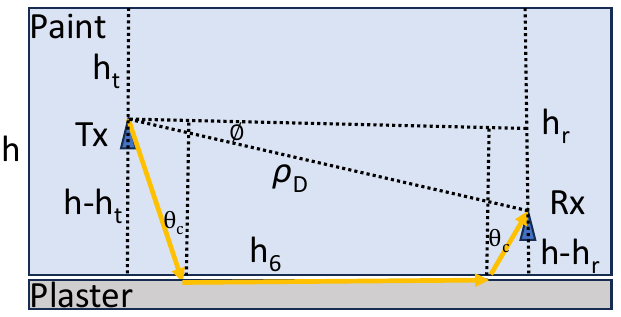}
    \caption{Geometry for the Lateral wave propagating through Paint-Plaster interface, where energy leakage back into the paint leads to a specific wave that aligns to the receiver.}
    \label{fig:Late_Paint_Plaster}
\end{figure}

Therefore, the received power when utilizing a lateral wave along the P-P interface is given as,
\begin{equation}
\begin{split}    
    P_r^{L,PP} (dBm) =& P_t (dBm) + G_t (dBi) + G_r (dBi) \\
    &- PL_{L,PP} (dB),
\end{split}
\end{equation}
where $P_t$ is the transmitted power, $G_t$ and $G_r$ are the transmitter and the receiver gains.

\subsection{Overall Received Power}
For an isotropic antenna, the overall received power can be expressed as the sum of the five components \cite{dong2011channel} as:
\begin{equation}
    \label{eq:Total_received_power}
    \begin{split}
        P_r^{o} = &10\log\left[10^{(P_r^{D}/10)} + 10^{(P_r^{R,AP}/10)} +10^{(P_r^{R,PP}/10)}\right.\\
        & \left. +10^{(P_r^{L,AP}/10)}
        + 10^{(P_r^{L,PP}/10)}\right].
\end{split}
\end{equation}

\section{Channel Capacity Model}
\label{sec:channel_capacity}
The capacity of the IoP channel at the THz frequencies can be obtained by partitioning the overall bandwidth into multiple narrow sub-bands and then aggregating the individual capacities because of its pronounced frequency selectivity and the existence of non-white molecular noise \cite{goldsmith2005wireless, jornet2010capacity, jornet2011channel}.

For a given operation band of $a~\text{to}~b$ in THz, the band can be divided into multiple sub-bands with bandwidth $\Delta f= (b-a)/n$, where $n$ is a positive integer selected corresponding to the required bandwidth of the sub-band. Moreover, the $i^{th}$ sub-band is centered around the frequency $f_i$ where $i=1, 2, ..., n$. Therefore, for a small enough sub-band, the noise power spectral density (p.s.d.) can be assumed flat in the entire THz band ($0.1-10$\,THz), and the resulting capacity for the $i^{th}$ sub-band in bits/s is expressed as \cite{jornet2010capacity}:
\begin{equation}
    C_i(d) = \Delta f \log_2{\left[ 1+ \dfrac{S(f_i)}{A(f_i,d)N(f_i,d)}\right]}, ~~~ i= 1, 2, ..., n
\end{equation}
where, $d$ is the total path length, $S$ is the transmitted signal power spectral density (PSD), $A$ is the channel total path loss, and $N$ is the noise PSD. Thus, the total channel capacity in the wide band can be expressed as the aggregated capacities of each narrow band and can be expressed as,
\begin{equation}
\label{eq_chanel_capacity}
    C(d) = \sum_i^n{C_i(d)}.
\end{equation}
The total channel path loss ($A(f_i,d)$) for each center frequency $f_i$ and for total propagation distance $d$ considering all five paths can be found utilizing (\ref{eq:Total_received_power}) as explained in the Section \ref{sec:channel_model}. The advancements in graphene-based nanoelectronics have enabled the transmission of incredibly short pulses, as short as one hundred femtoseconds \cite{da2009carbon}. The capability to produce such power is provided within the THz spectrum. Furthermore, short-pulse signals are attributed to the emergence of lateral waves on the boundaries of each media~\cite{King92Book,Wabia92}. Nano-network devices can also transmit very short picosecond long pulses in the narrow THz frequency bands, and they can be modeled as Gaussian-shaped \cite{jornet2010capacity}:
\begin{equation}
    p(t) = \dfrac{a_0}{\sqrt{2\pi}\sigma}e^{-(t-\mu)^2/2\sigma^2},
\end{equation}
where $a_0$ is a normalizing constant to adjust the pulse total energy (500~pJ), $\sigma \in (0.005~\text{ps}, 0.15 ~\text{ps})$ is the standard deviation of the Gaussian pulse in seconds, $\mu$ is the location in time for the center of the pulse in seconds. Therefore, the PSD of the transmitted pulse can be expressed as, 
\begin{equation}
\label{eq_psd_transmitted}
    S(f_i) = a_0^2e^{-(2\pi \sigma f_i)^2}, ~~~ i= 1, 2, ..., n.
\end{equation}

Corresponding to our model, the noise in the THz channel is mainly contributed by the absorption noise when propagating through a specific medium. The emissivity parameter, which measures this noise phenomenon of the
channel is defined as,
\begin{equation}
    \epsilon(f_i,d) = 1 - \tau_p(f_i,d), ~~~ i= 1, 2, ..., n
\end{equation}
where $\tau_p$ is the transmittance of the paint medium, where the nano-devices reside. The transmittance measures the fraction of incident radiation that is able to pass through the medium, and it can be expressed using Beer-Lambert Law as \cite{Beer-Lam_1995}:
\begin{equation}
    \tau(f_i,d) = e^{-K_p(f_i)d}, ~~~ i= 1, 2, ..., n
\end{equation}
where $K_p(f_i)$ is the paint medium absorption coefficient at frequency $f_i$ as given in (\ref{eq:kf-paint}). The equivalent noise temperature due to medium absorption $T_\textsubscript{mol}$ (in Kelvin) that an antenna will detect from the medium is obtained as \cite{jornet2011channel}:
\begin{equation}
    T_\textsubscript{mol}(f_i,d)=T_0 \epsilon(f_i,d), ~~~ i= 1, 2, ..., n
\end{equation}
where $T_0$ (296~K) is the reference temperature. Therefore, for a given bandwidth, the noise PSD is given by
\begin{equation}
\label{eq_psd_noise}
    N(f_i,d) = \Delta{f} k_B T_\textsubscript{mol}(f_i,d), ~~~ i= 1, 2, ..., n
\end{equation}
where $k_B$ is the Boltzmann constant. Therefore, substituting the PSD expression obtained for the transmitted pulse (\ref{eq_psd_transmitted}) and noise (\ref{eq_psd_noise}) into the derived channel capacity expression (\ref{eq_chanel_capacity}), we can rewrite the channel capacity expression for a wide-band as:
\begin{equation}
    C(d) = \sum_i^n{\Delta f \log_2{\left[ 1+ \dfrac{a_0^2e^{-(2\pi \sigma f_i)^2} }{\Delta{f} k_B T_\textsubscript{mol}(f_i,d)A(f_i,d)}\right]}}.
\end{equation}

\begin{table}[t]
    \centering
    \begin{tabular}{|c|c|c|}
    \hline
       Paint type  & Refractive index ($n_p$) & Main pigment\\
       \hline
      Brilliant Blue   & 1.91 &PB15\\
      Titanium White   & 2.13 &PW6\\
      Oxide Black   & 2.74 &PBk11\\
\hline
    \end{tabular}
    \caption{Types of paint and corresponding mean refractive index ($\pm0.05$) in the frequency
range 0.1 to 1.2\,THz \cite{abraham2010non}.}
    \label{tab:Paint_type}
\end{table}

\section{Results and Discussion}
\label{sec:Results_and_Discussion}
In this section, we perform numerical evaluations of the path loss and channel capacity of the IoP channels in the frequency range of $200$ to $300$\,GHz, using MATLAB, following the methodology explained in Section \ref{sec:channel_model} and \ref{sec:channel_capacity}. We assume that technology is available to embed transceivers in a paint layer with a thickness of $h$. 
The transmitted power of each proposed nano-network device is considered to be $10$\,mW ($10$\,dBm), with $2.15$\,dBi gains for both the transmitter and the receiver \cite{elayan2017terahertz}. To analyze the impacts of paint color, we consider three types of paints: Brilliant Blue, Titanium White, and Oxide Black (Table \ref{tab:Paint_type}). The paint is applied to a plaster wall with a refractive index ($n$) of $1.61$ \cite{refractive2008}. The refractive index of the paint types and plaster can be considered constant with frequency since the variation is negligible in the frequency range considered in this study \cite{refractive2008}.
We consider the thickness $h$ of the paint to vary between $1$\,mm-$3$\,mm \cite{otoshi1999measurements}, and the minimum distance between the transmitter and the receiver is set at $4$\,mm. 

\subsection{Path Loss Analysis}
\label{subsec:PL_and_RP}

We first analyze the path loss for all the critical communication paths based on the analysis in Section \ref{sec:channel_model}. Assuming a paint layer thickness of $h$ = $2$\,mm, we consider the burial depths of the transmitter and the receiver antennas at $0.5$\,mm and $1.5$\,mm from the A-P interface, respectively. The path loss is analyzed by varying the frequency ($200-300$\,GHz) as well as the distance between the transmitter and the receiver ($4$\,mm to $10$\,cm), with default values of $200$\,GHz and $1$\,cm, respectively. The absorption coefficients for both the paint and plaster are obtained from \cite{refractive2008}. The molecular absorption coefficient was calculated by considering all the gases in Table \ref{tab:AtmosphereComparison} except \ce{Ar}, as explained in section \ref{subsec:LW_A_P}.

\subsubsection{Direct Wave}

\begin{figure}[t]
    \centering
    \subfigure[]{\includegraphics[width=0.24\textwidth]{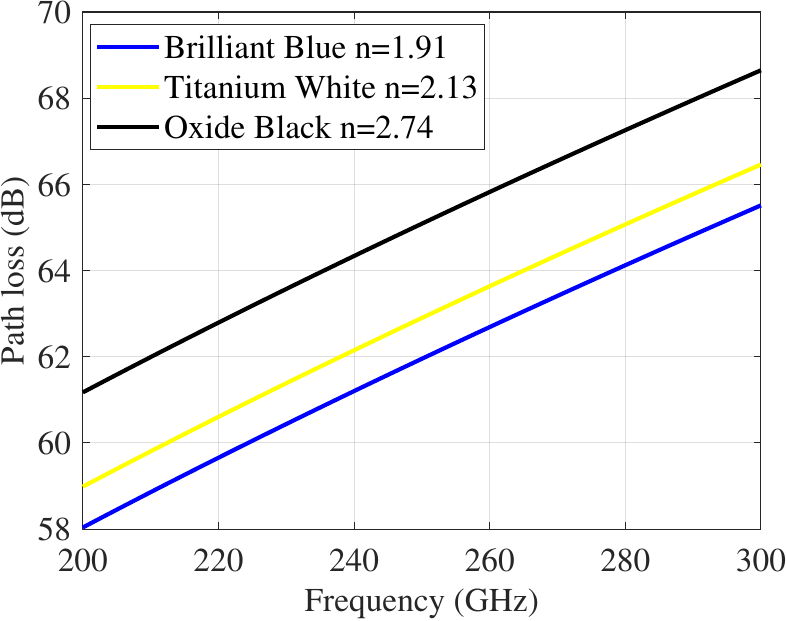}} 
    \subfigure[]{\includegraphics[width=0.24\textwidth]{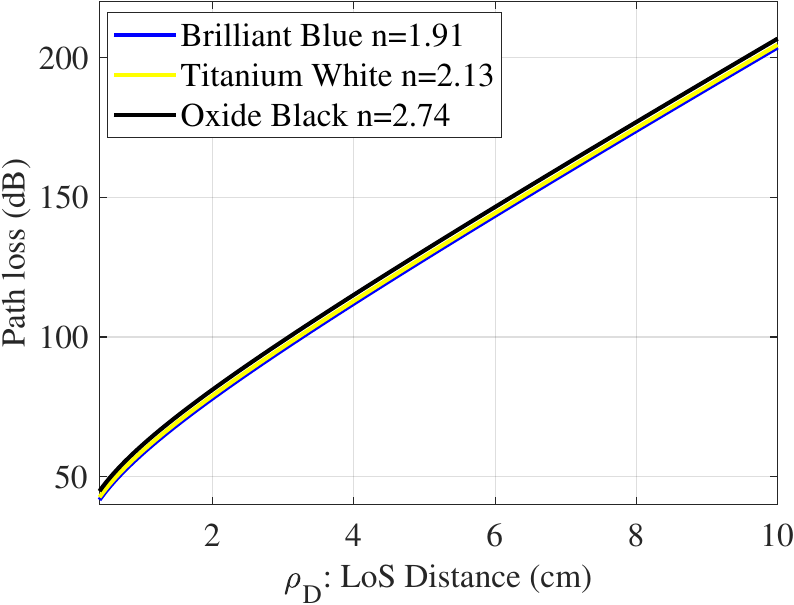}} 
    \caption{Path loss of the direct wave as a function of (a) frequency (fixed $\rho_D$ = $1$\,cm) and (b) distance (fixed f= $200$\,GHz) for three paint colors. The burial depth of the transmitter and the receiver antennas are $0.5$\,mm and $1.5$\,mm respectively}.
    \label{fig:DW_PL_RP_dis_fre}
\end{figure} 
 
The path loss of the direct wave is shown in Figs.~\ref{fig:DW_PL_RP_dis_fre} for the three different paint pigments with respect to a range of frequencies as well as distances between the transmitter and the receiver. When varying the frequency from $200$\,GHz to $300$\,GHz, the path loss increases gradually at a rate of $0.075$\,dB/GHz for all paint colors. The path loss is higher for paint types with higher refractive indices, where a $3.14$\,dB increase in path loss is observed between blue and black paint in Fig.~\ref{fig:DW_PL_RP_dis_fre} (a). The path loss increases linearly due to the corresponding increase in the absorption coefficient of the paint in the $200$\,GHz to $300$\,GHz frequency range \cite{refractive2008}.
 
In Fig.~\ref{fig:DW_PL_RP_dis_fre} (b), the impact of distance on path loss is shown for three paint colors. It is important to note that the spreading loss dominates the path loss in the first few centimeters and a further increase in distance lead to higher absorption loss. This behavior explains the exponential increase in the path loss initially, followed by a moderate linear increase at a rate of $15.28$ dB/cm.

\subsubsection{Reflected Wave from A-P interface}

\begin{figure}[t]
    \centering
    \subfigure[]{\includegraphics[width=0.24\textwidth]{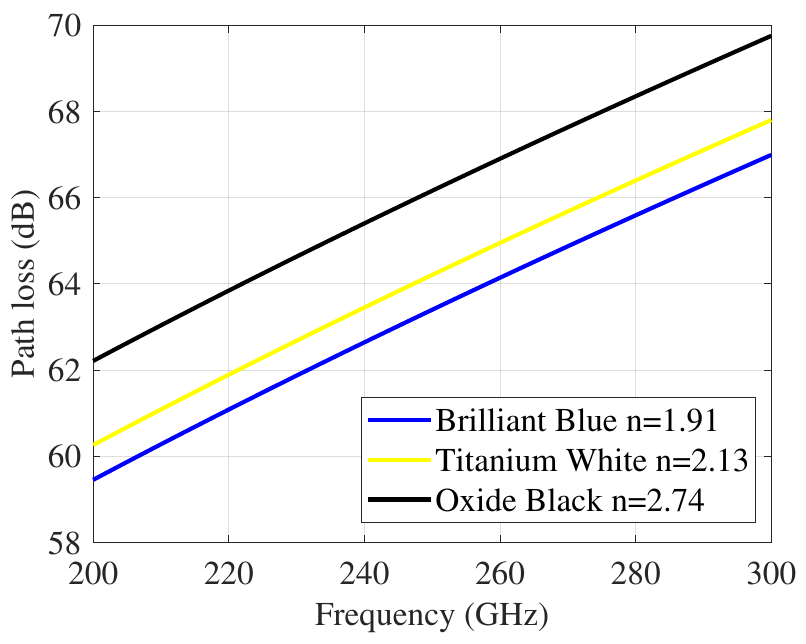}} 
    \subfigure[]{\includegraphics[width=0.24\textwidth]{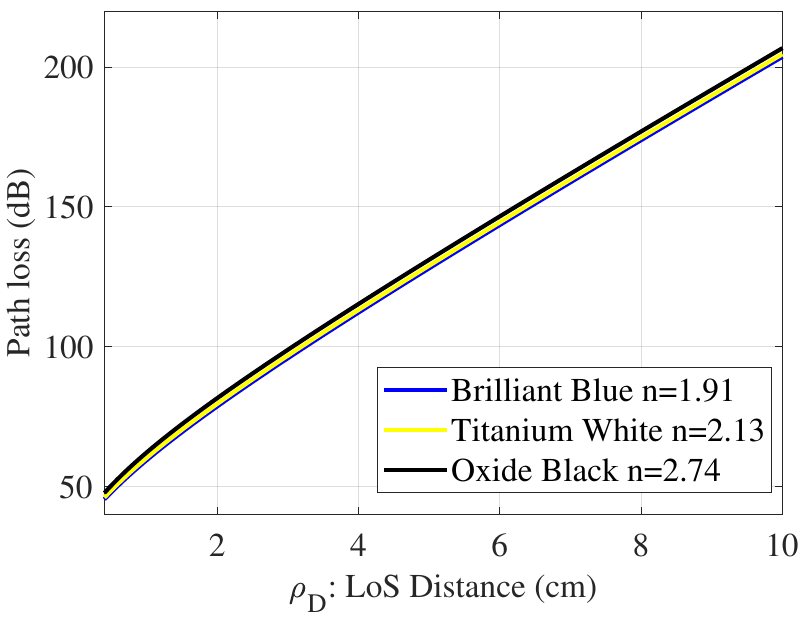}} 
    \caption{Path loss of the reflected wave from the A-P interface as a function of (a) frequency (fixed $\rho_D$ = $1$\,cm) and (b) distance (fixed f= $200$\,GHz) for three paint colors. The burial depth of the transmitter and the receiver antennas are $0.5$\,mm and $1.5$\,mm respectively.}
    \label{fig:RW_AP_PL_RP_dis_fre}
\end{figure}

The RW-A path loss is calculated according to (\ref{eqn:reflectedPL}) with a surface height standard deviation of $0.01$\,mm~\cite{refractive2008}. The results are shown in Figs.~\ref{fig:RW_AP_PL_RP_dis_fre} with respect to varying frequencies and distances between the transmitter and the receiver. According to Fig.~\ref{fig:RW_AP_PL_RP_dis_fre}~(a), the path loss increases non-linearly with frequency with a rate of change around $0.07-0.083$\,dB/GHz, unlike the DW case. However, similar to the DW case, we notice clear differences in the total path losses for different paint pigments. Paints with higher refractive index have comparably higher path losses. However, RW-A experiences path loss $1$-$2$\,dB higher than the DW, because of the additional spreading and absorption losses that accrue along the longer reflected path ($h_1+h_2 > \rho_D$) as well as the reflection loss, $R_p(f)$. The differences in total path loss between Blue and White, and between White and Black pigments are $0.9$\,dB and $2.2$\,dB, respectively. This aligns with the differences in their refractive indexes. 

\subsubsection{Reflected wave from P-P interface}

\begin{figure}[t]
    \centering
    \subfigure[]{\includegraphics[width=0.24\textwidth]{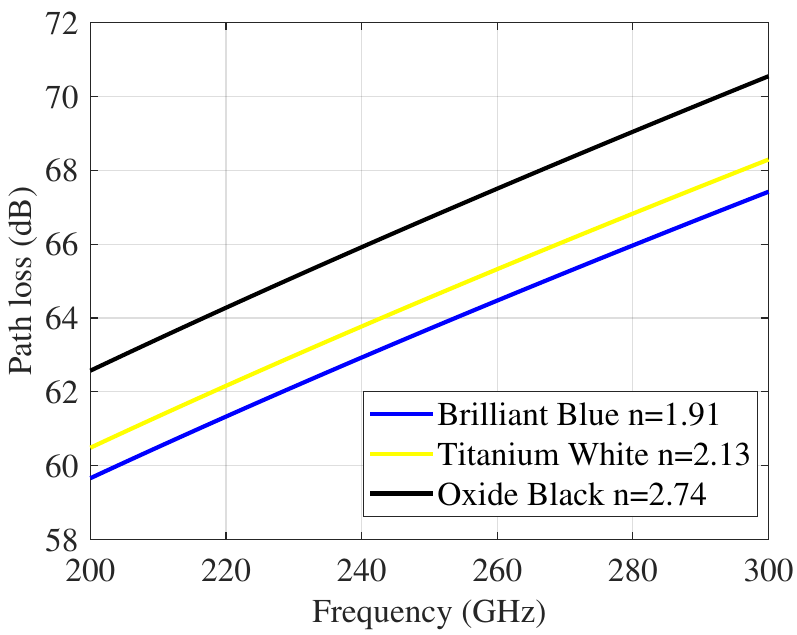}} 
    \subfigure[]{\includegraphics[width=0.24\textwidth]{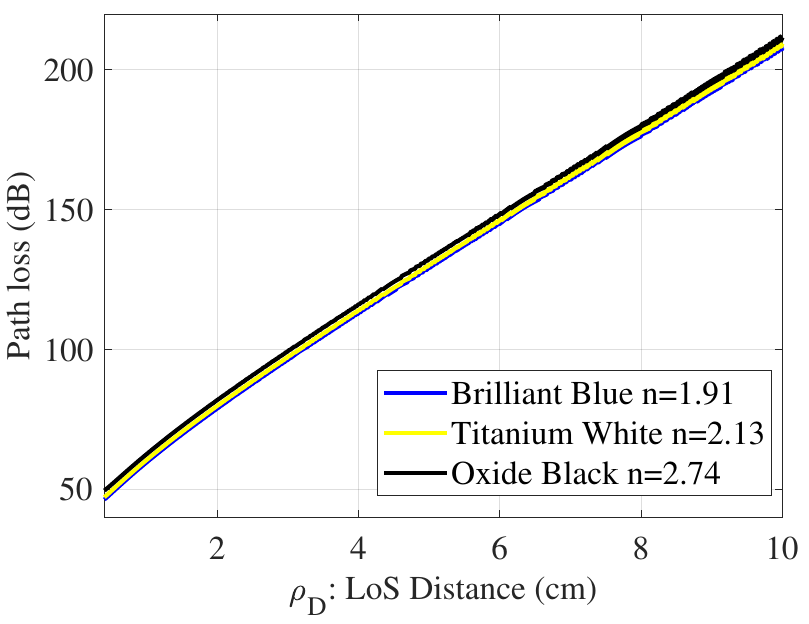}} 
    \caption{Path loss of the reflected wave from the P-P interface as a function of (a) frequency (fixed $\rho_D$ = $1$\,cm) and (b) distance (fixed f= $200$\,GHz) for three paint colors. The burial depth of the transmitter and the receiver antennas are $0.5$\,mm and $1.5$\,mm respectively.}
    \label{fig:RW_PP_PL_RP_dis_fre}
\end{figure}

The path loss of the RW-P is analyzed similarly to RW-A, with a major difference in that the rough surface height standard deviation, which is $\sigma_{pl}=0.05$\,mm at the P-P interface is expected to be larger than that at the A-P interface ($\sigma_p=0.01$\,mm) \cite{piesiewicz2007scattering}. As shown in Figs.~\ref{fig:RW_PP_PL_RP_dis_fre}, RW-P behaves similar to RW-A (Fig.~\ref{fig:RW_AP_PL_RP_dis_fre}). 
The main difference is that the path loss increases at a slightly higher rate with the frequency ($0.0043$\,dB/GHz) and the LoS distance ($0.5528$\,dB/cm) than the corresponding A-P results. This is mainly caused by the greater reflection loss and roughness at the P-P interface compared to the A-P interface.

\begin{figure}[t]
    \centering
    \subfigure[LW-A.]{\includegraphics[width=0.24\textwidth]{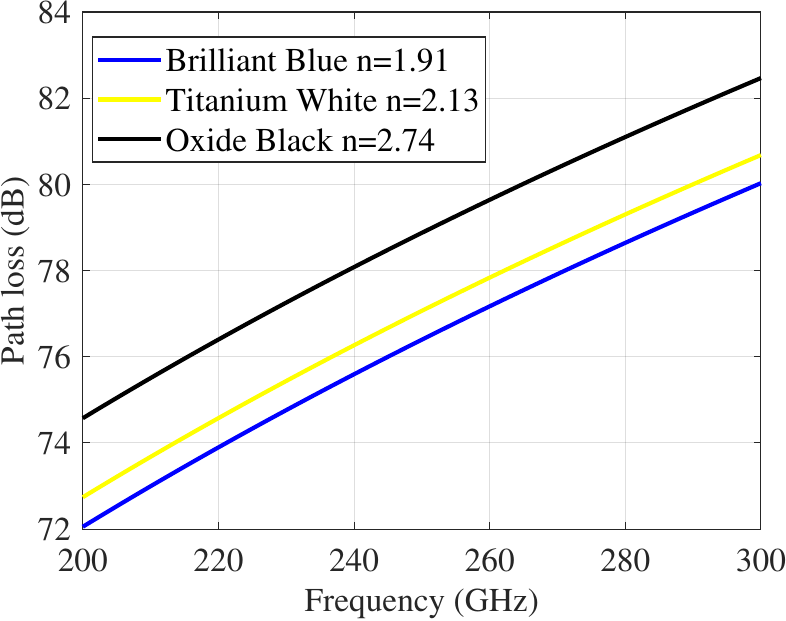}} 
    \subfigure[LW-P.]{\includegraphics[width=0.24\textwidth]{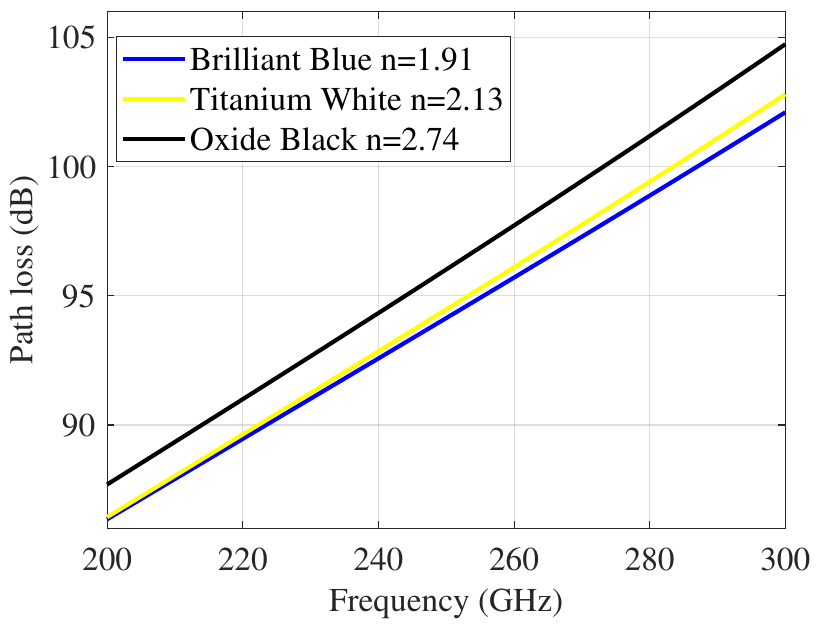}}
    \caption{Path loss of the lateral waves through (a) A-P and (b) P-P interfaces as a function of frequency (fixed $\rho_D$ = $1$\,cm) for three paint colors. The burial depth of the transmitter and the receiver antennas are $0.5$\,mm and $1.5$\,mm respectively.}
    \label{fig:LW_AP_PP_PL_RP_fre}
\end{figure} 

\begin{figure}[t]
    \centering
    \subfigure[LW-A.]{\includegraphics[width=0.24\textwidth]{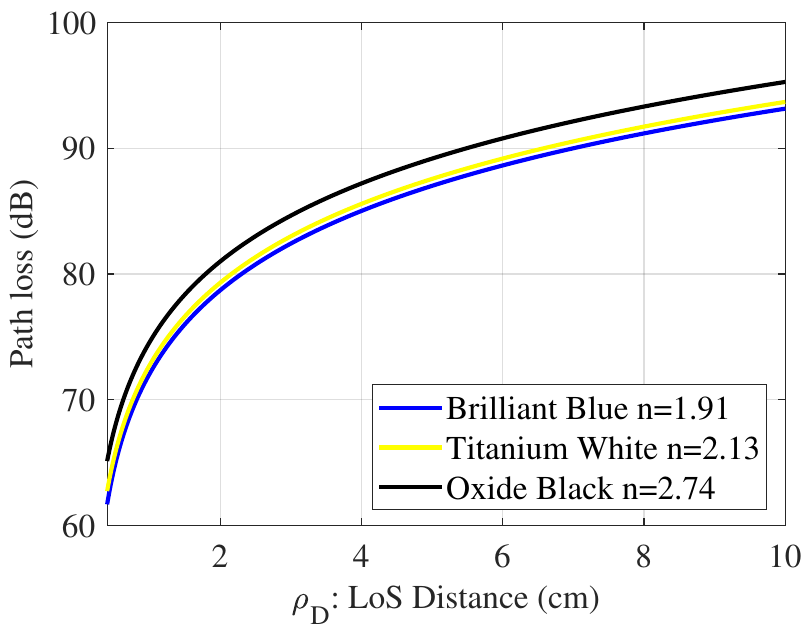}} 
    \subfigure[LW-P.]{\includegraphics[width=0.24\textwidth]{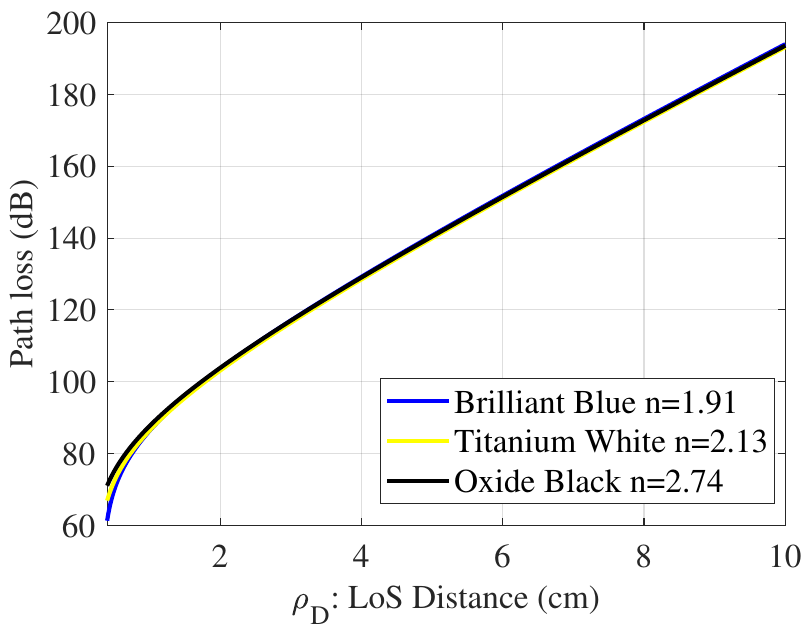}} 
    \caption{Path loss of the lateral waves through (a) A-P and (b) P-P interfaces as a function of distance (fixed f= $200$\,GHz) for three paint colors. The burial depth of the transmitter and the receiver antennas are $0.5$\,mm and $1.5$\,mm respectively.}
    \label{fig:LW_AP_PP_PL_RP_dis}
\end{figure} 

\subsubsection{Lateral wave through Air-Paint and Paint-Plaster interfaces}

As described in Section~\ref{sec:channel_model}, lateral waves are unique components of the IoP channel that are exhibited when signals from a medium with higher permittivity are incident to a boundary with a medium with lower permittivity at the critical angle of reflection according to Snell's Law. 

In Figs.~\ref{fig:LW_AP_PP_PL_RP_fre}, the lateral wave path loss as a function of frequency is shown for both A-P and P-P interfaces. It can be observed that the path loss is moderately sublinear with frequency in the case of the A-P interface, and moderately superlinear in the case of the P-P interface. Moreover, at the $1$ cm inter-node distance, the path loss for lateral waves is significantly larger than that of direct and reflected waves. More specifically, at $200$\,THz and for blue paint, path losses for LW-A and LW-P are 72\,dB and 86\,dB, respectively. By contrast, the corresponding path losses direct and reflected waves are $58$\, and $60$\,dB, respectively. However, as we will analyze in Section \ref{subsec:Burial_Depth_Analysis}, lateral waves start to dominate (i.e., result in lower path loss) other waves as the burial depth and inter-node distance are changed. Moreover, it can be observed that the LW-P and LW-A path loss difference corresponding to each frequency is increasing from $13$ to $22$\,dB. This is mainly because of the higher reflective index of plaster compared to that of air as well as the changes in the paint paths as waves propagate to these respective interfaces. 

In Fig. \ref{fig:LW_AP_PP_PL_RP_dis} (a), the LW-A path loss is shown as a function of distance, $\rho_D$. A unique property of lateral waves is observed in the results. More specifically, the path loss is not linearly increased with distance compared to direct and reflected waves (see Figs.~\ref{fig:DW_PL_RP_dis_fre} (a), \ref{fig:RW_AP_PL_RP_dis_fre} (a), and \ref{fig:RW_PP_PL_RP_dis_fre} (a)). This is mainly due to the significantly lower permittivity of air and hence the lower path loss as the majority of lateral wave propagates through air. As a result, as the distance between the transmitter and receiver increases, the lateral wave dominates the channel, maintaining a higher communication range compared to other waves. In Fig. \ref{fig:LW_AP_PP_PL_RP_dis} (b), the LW-P path loss is shown, where the path loss values are $13$--$22.3$ dB higher than that of LW-A. This increase is mainly due to the higher path loss within the plaster compared to air. 

\subsection{Burial Depth Analysis}
\label{subsec:Burial_Depth_Analysis}

\begin{figure*}[t]
    \centering
    \subfigure[Paint Thickness h = 1\,mm.]{\includegraphics[width=0.3\textwidth]{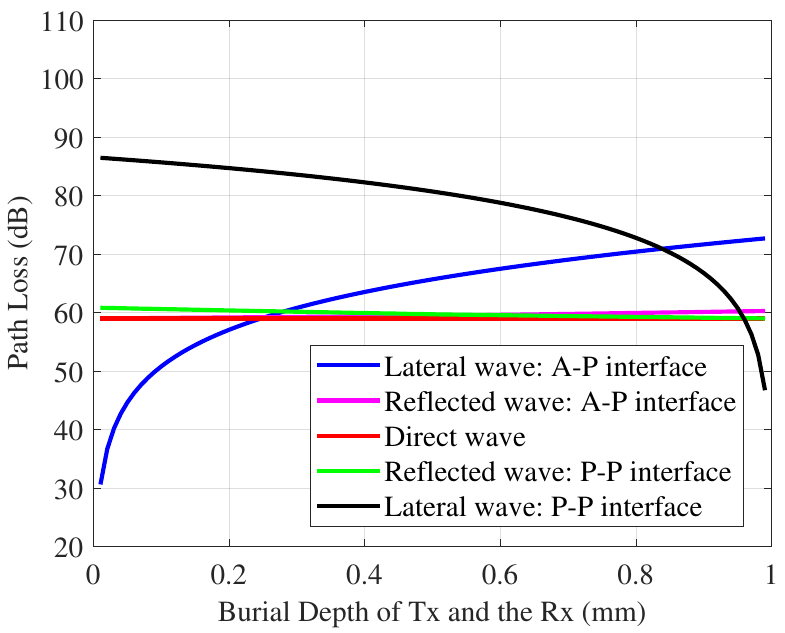}} 
    \subfigure[Paint Thickness h = 2\,mm.]{\includegraphics[width=0.3\textwidth]{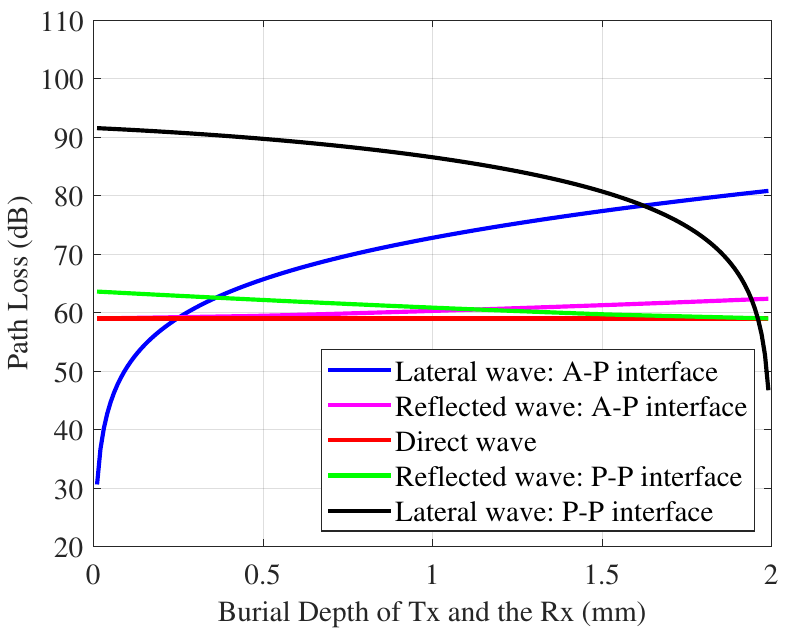}}
    \subfigure[Paint Thickness h = 3\,mm.]{\includegraphics[width=0.3\textwidth]{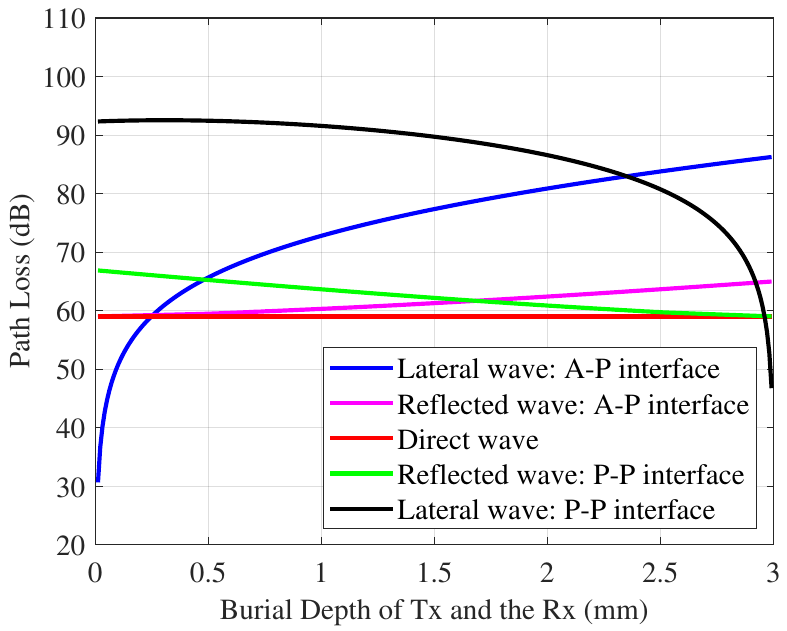}} 
    \caption{Path loss variation with the burial depth for the transmitter and the receiver for a fixed LoS distance (1\,cm) and frequency (200\,GHz) for Titanium White paint (n = 2.13).}
    \label{fig:pathloss_vs_burial_depth}
\end{figure*} 

\begin{figure*}[t]
    \centering
    \subfigure[LoS Distance $\rho_D$ = 2\,cm.]{\includegraphics[width=0.3\textwidth]{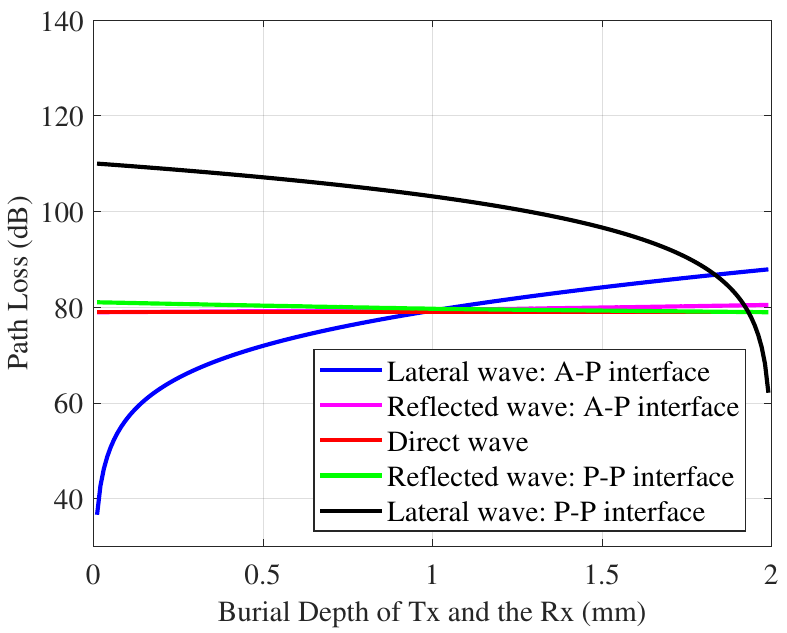}}
    \subfigure[LoS Distance $\rho_D$ = 3\,cm.]{\includegraphics[width=0.3\textwidth]{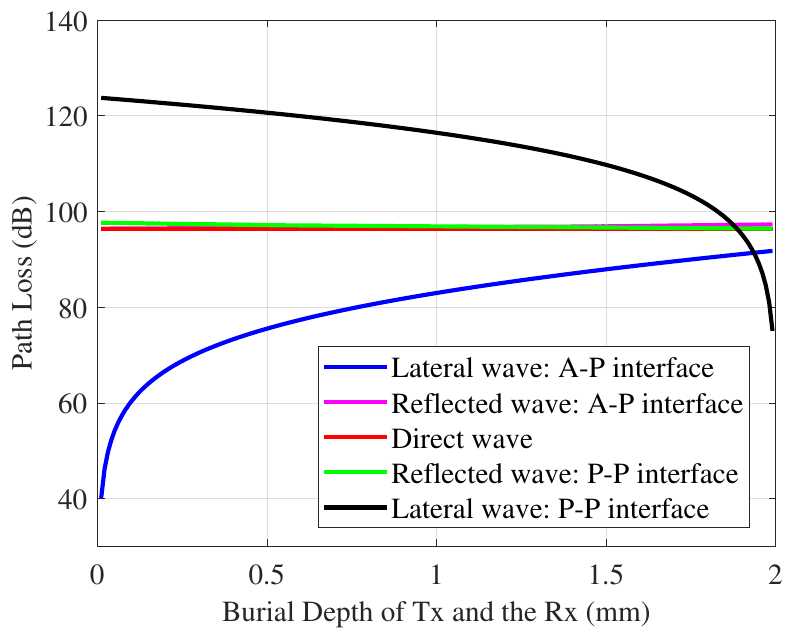}}
    \subfigure[LoS Distance $\rho_D$ = 4\,cm.]{\includegraphics[width=0.3\textwidth]{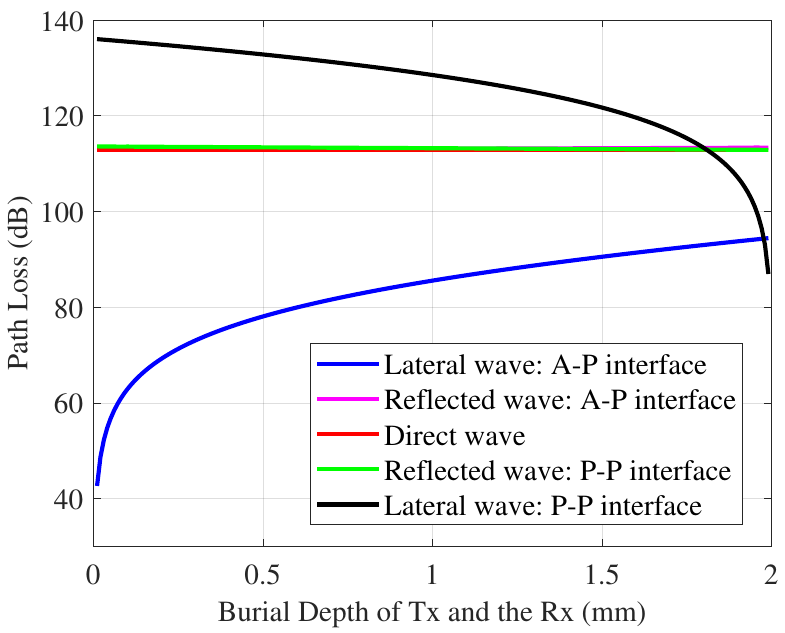}}
    \caption{Path loss variation with the burial depth for different LoS distances (2 to 4\,cm) for a fixed frequency (200\,GHz) and thickness (2\,mm) for Titanium White paint (n = 2.13) (LoS distance of $1$\,cm can be seen in Fig.~\ref{fig:pathloss_vs_burial_depth} (b)).}
    \label{fig:pathloss_vs_LoS_dis}
\end{figure*}

So far, we have analyzed the impacts of frequency and distance on IoP channel by fixing the burial depths of the transmitter and receiver. However, as we discussed above, burial depth governs the reflected and lateral waves because the signal path in paint varies accordingly. Next, we investigate the performance of the direct, reflected, and lateral waves by varying the burial depths of the transmitter and the receiver, assuming that the technology is available to keep the transceivers at the same level. The burial depth is measured from the A-P interface. In addition, we vary the paint thickness ($1$-$3$\,mm) to explore the impacts of burial depth on different paint applications. 

\begin{figure}
    \centering
    \includegraphics[width=\linewidth]{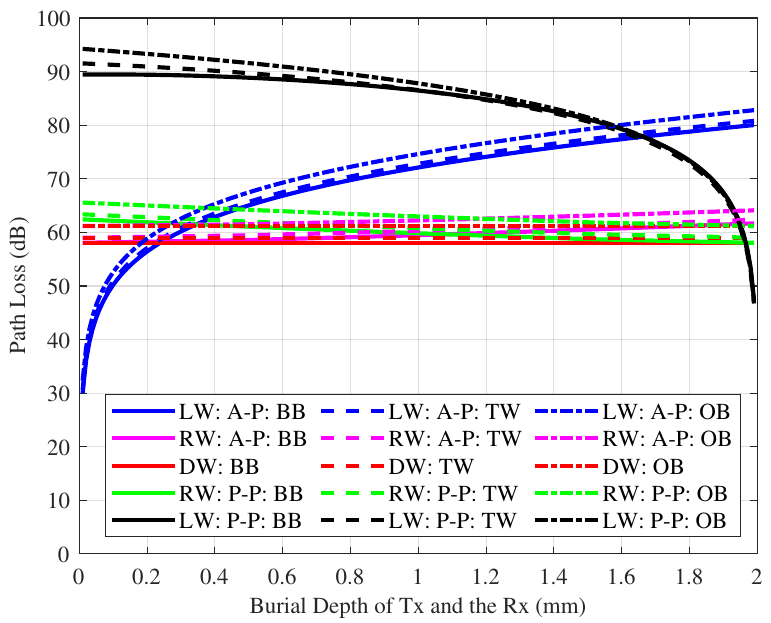}
    \caption{Path loss variation with the burial depth of the transceivers for fixed frequency (200\,GHz), LoS distance (1\,cm), and thickness (2\,mm) for all paint types. BB - Brilliant Blue, TW - Titanium White, OB - Oxide Black.}
    \label{fig:Pathloss_all_paint}
\end{figure}

\begin{figure*}[t]
    \centering
    \subfigure[Paint Thickness h = 1\,mm.]{\includegraphics[width=0.3\textwidth]{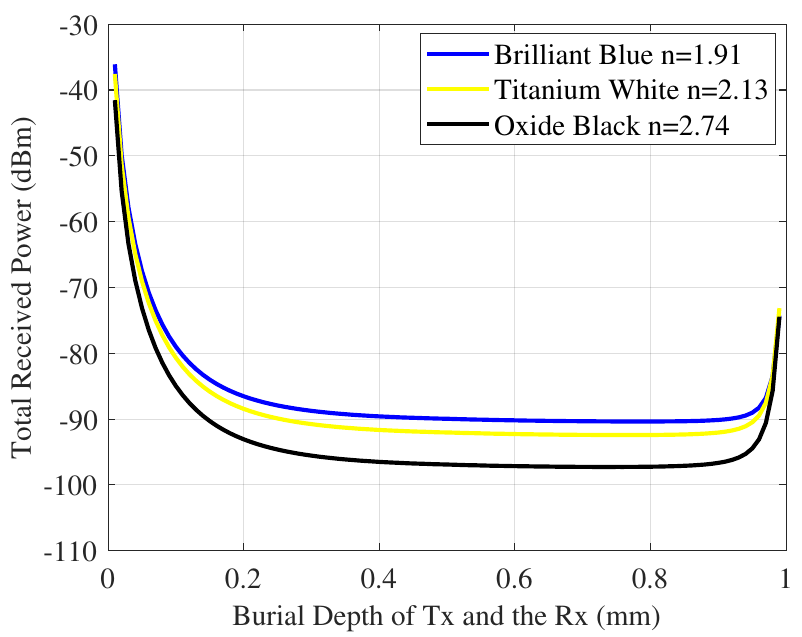}} 
    \subfigure[Paint Thickness h = 2\,mm.]{\includegraphics[width=0.3\textwidth]{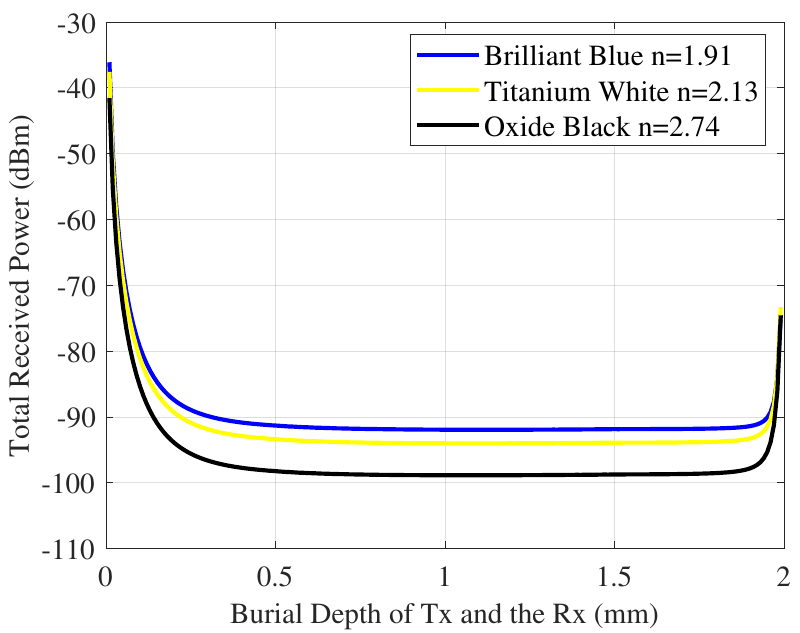}}
    \subfigure[Paint Thickness h = 3\,mm.]{\includegraphics[width=0.3\textwidth]{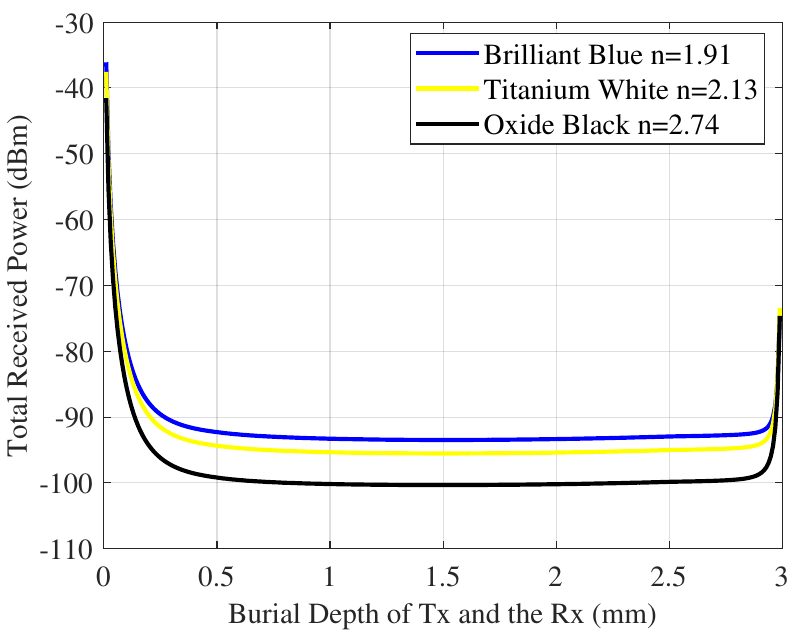}} 
    \caption{Total received power variation with the burial depth for a fixed distance between the transmitter and the receiver (1\,cm) and frequency (200\,GHz) for all paint types.}
    \label{fig:TRP_vs_B_depth}
\end{figure*}

In Figs.~\ref{fig:pathloss_vs_burial_depth}, the path loss values of the five main paths are shown as a function of burial depth for three paint thickness values. The results in Figs.~\ref{fig:pathloss_vs_burial_depth} clearly illustrate the dynamics between each path as the burial depth and the paint thickness changes. For example, it can be observed that DW is independent of the burial depth and the paint thickness, as expected. However, the other types of waves are significantly affected by the thickness of the paint due to the increase in wave propagation distance. The lateral waves propagating through the A-P and P-P interfaces are the most reliable communication paths (i.e., lowest path loss) when the transceivers are in close proximity to the corresponding interfaces. In all other scenarios, for the fixed distance of $1$ cm, the DW results in the best performance for all paint thickness values. When analyzing the path loss corresponding to the reflected waves, we can observe that there is a burial depth value at which the path loss of the RW-A interface gets higher than the RW-P interface, shifting towards the P-P interface as the thickness of the paint increases. The reason for this behaviour is due to the increased surface roughness of the plaster compared to the paint. 
Furthermore, we can observe that the path loss associated with the reflected waves from the A-P and P-P interfaces is almost the same as the DW at a 1\,mm thickness. This is due to the fact that the path lengths within the thin paint layer are nearly equal. Thus, with the increase in thickness, the path losses corresponding to the reflected waves exhibit significant differences.   

Since the transceiver could be positioned inside the paint at various distances, it would be imprecise if we made decisions on the communication through the paint for a constant LoS distance. Therefore, in Figs. \ref{fig:pathloss_vs_LoS_dis}, we present the path loss variation against the burial depths of the transceivers by varying the LoS distance from $1$ to $4$\,cm (path loss for $1$\,cm can be observed in Fig.~\ref{fig:pathloss_vs_burial_depth} (c)) for fixed paint thickness of $2$\,mm. As the distance increases, the path loss values corresponding to all the types of waves increase. This is due to the fact that absorption and spreading losses are directly proportional to the distance. However, it can also be observed that the rate of increase in path loss with distance is significantly smaller for lateral waves through the A-P interface, where LW-A dominates the IoP channel for long distances. For example, at a burial depth of $1$\,mm, an increase in distance from $1$\,cm to $4$\,cm leads to only a $12.79$\,dB increase in the LW-A path loss, compared to an increase of $53.89$\,dB, $52.82$\,dB, $52.39$\,dB, and $41.97$\,dB in path loss for DW, RW-A, RW-P and LW-P, respectively. Moreover, we observe that LW-P dominates the IoP channel only when transceivers are within $0.08$\,mm of the P-P interface. The reason for this behavior is the significantly higher absorption properties of the plaster than the molecular absorption in air. 

Moreover, it is necessary to investigate the impact of different paint types on communication through paint because different paint types have different refractive indices. In Fig.~\ref{fig:Pathloss_all_paint}, the path loss values of the five main paths are shown for three paint colors, for a fixed distance of $1$\,cm, paint thickness of $2$\,mm, and a frequency of $200$\,GHz. It can be observed that the path loss values corresponding to the paint types with higher refractive index lower the performance.

From the practical operation of IoP at sub-THz frequencies, two salient observations emerge. Note that in this study, we explore every major IoP communication path. However, in practice, the use of directional antennas can selectively enhance or merge specific paths to improve communication efficacy. For communication distances below $2$\,cm, multiple paths experience similar path loss (e.g., DW, RW-A, RW-P). This suggests that leveraging multi-beam transmission paired with receiver combining could offer significant improvements in range and channel performance. On the other hand, for larger communication distances, lateral waves through the A-P dominate the channel. To maximize efficiency in such scenarios, sub-THz waves could be steered towards the critical angle, concentrating energy to the lateral waves and thus conserving overall energy consumption. Interestingly, due to the difference in refractive indices of paint colors, such IoP steering and multi-beam solutions would need to be designed based on paint color. 

\begin{figure}[t]
    \centering
    \includegraphics[width=\linewidth]{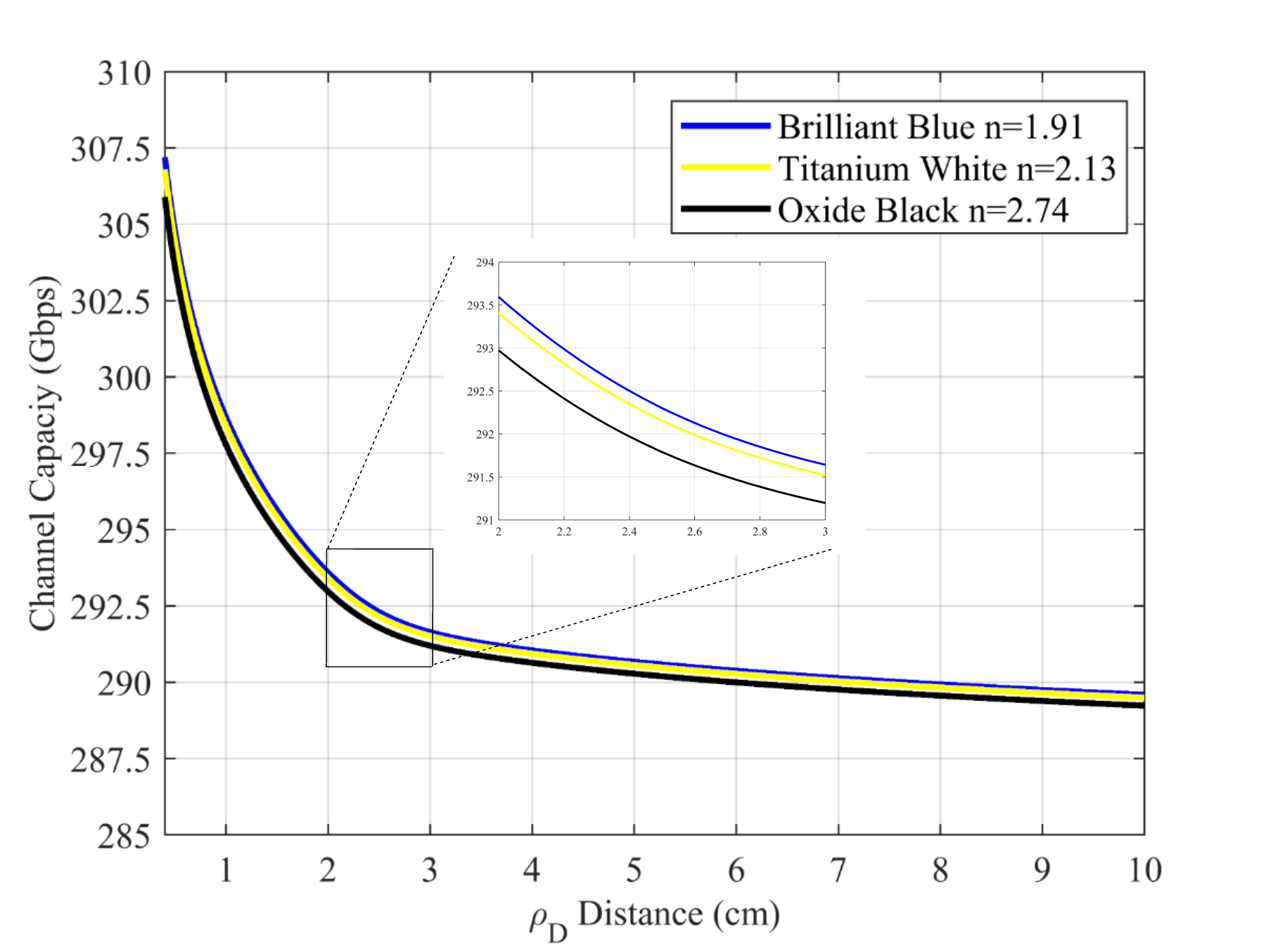}
    \caption{Multi path Channel capacity in 200 to 300\,GHz frequency band for all paint types.}
    \label{fig:CC_all}
\end{figure}

Finally, combining all the received power corresponding to each individual communication path (see Fig.~\ref{fig:TRP_vs_B_depth}), we investigate the total received power variation with the burial depth of the transceivers by keeping the frequency of $200$\,GHz and LoS distance of $1$\,cm constant. It can be observed that there is a $40$-$50$\,dB improvement in received power when the transceivers are located close to the A-P interface. As the burial depth is increased, the total received power decays exponentially with the burial depth of the transceivers at the first few microns (approximately $h_t < 0.3\,mm$ to $0.5$\,mm) close to the A-P interface, stabilizing in the middle range (approximately $0.3$\,mm $< h_t < 0.9$\,mm). Finally, a considerable increase ($10$-$15$\,dB) is observed when transceivers are close to the P-P interface ($h_t>0.9$\,mm). This behavior can be seen in all the paint types and thicknesses due to the reduced path loss when the transceivers are close to the A-P and P-P interfaces. Up to 33\,dB improvement in received power can be achieved by placing transceivers close to the A-P interface compared to the P-P interface. Moreover, paints with higher refractive indices absorb more signal energy, reducing the received power compared to paints with lower refractive indices by up to 6.8\,dB. Furthermore, as the thickness of the paint layer increases, a $1$-$3.5$\,dB drop in total received power is observed, even though a considerable path loss increase is observed for each individual path, as shown in Fig. \ref{fig:pathloss_vs_burial_depth}. The results highlight the importance of multi-path communication and receiver combining in IoP as well as the importance of the placement process to improve channel quality and range.

 \begin{figure}[t]
    \centering
    \subfigure[]{\includegraphics[width=0.45\textwidth]{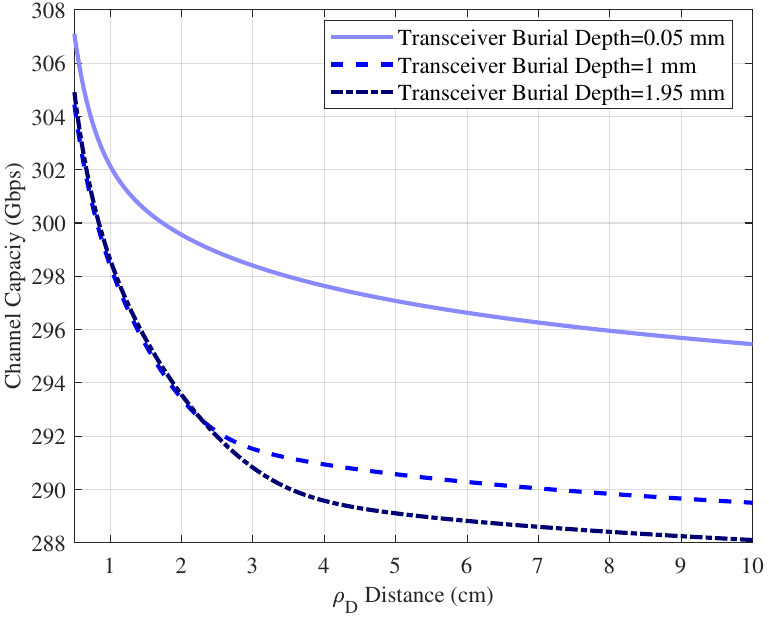}} 
    \subfigure[]{\includegraphics[width=0.45\textwidth]{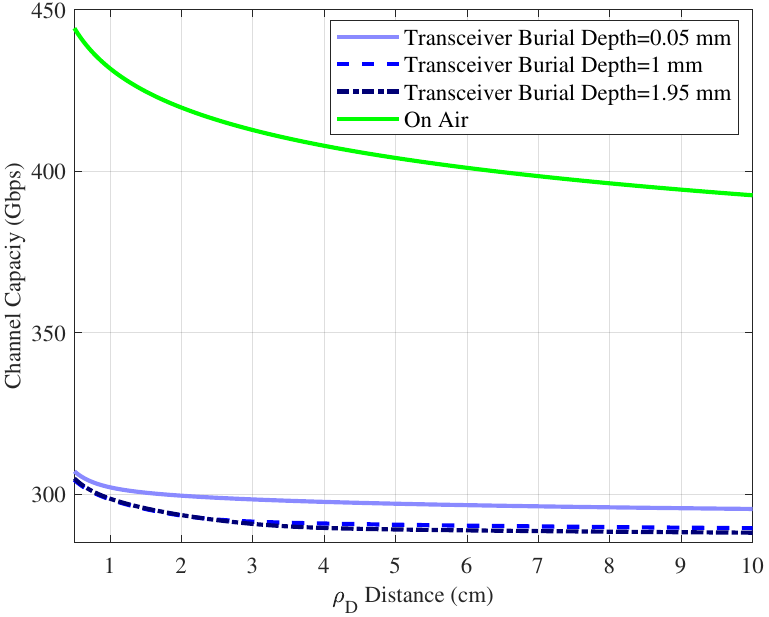}} 
    \caption{Comparison of Channel capacity through Air and Paint (Multipath) for various transceiver burial depths within $200$ to $300$\,GHz frequency. The thickness is considered as $2$\,mm for Titanium White paint (n = 2.13).}
    \label{fig:CC_Air_paint}
\end{figure}

\subsection{Channel Capacity Analysis}
\label{subsec:Channel_Capacity_Analysis}
The IoP channel capacity is analyzed based on the discussion in Section \ref{sec:channel_capacity}, by equally partitioning the frequency bands of $200-300$\,GHz into narrow bands with a bandwidth of $10$\,GHz, while increasing the LoS distance up to $10$\,cm between the transceivers. The computed channel capacity values for multipath communication through paint are presented in Fig.~\ref{fig:CC_all} for all the paint colors. Similar to the path loss analysis, as expected from the previous analysis, we can achieve higher channel capacities by applying paints with lower refractive indexes, which have lower losses. When investigating the channel capacity for the LoS distances between $2$ to $3$\,cm, we notice $0.1$ to $0.2$\,Gbps and $0.3$ to $0.45$\,Gbps difference between Blue and White, and between White and Black paints, respectively, and these gaps decrease slightly with distance. We do not recommend establishing communication between nanodevices more than $5$\,cm apart due to the exponential decrease in channel capacity with distance and thickness of the paint layer varying significantly. However, due to the random spatial distribution of the devices, the LoS distance to the nearest device is highly variable and not easily controlled. Therefore, to ensure relatively short propagation distances, we propose creating a dense network in the paint layer with more nanodevices per unit volume. 

Lastly, we analyze IoP channel capacity by varying the transceiver burial depths and comparing them against the channel capacity in air, which comprises water vapor and a selection of gases from Table \ref{tab:AtmosphereComparison}. The thickness of the paint layer is considered as $2$\,mm. Therefore the selected burial depths indicate the channel capacity when the transceivers are buried 1) near the A-P interface (depth = $0.05$\,mm), 2) in the middle (depth = $1$\,mm), 
and 3) near the P-P interface (depth = $1.95$\,mm). Referring to the predicted channel capacity as a function of distance, for each burial depth in Fig. \ref{fig:CC_Air_paint} (a), a dramatic decrease of $15.8-17.9$\,Gbps in channel capacity can be noticed until $4$\,cm LoS Distance. However, when the transceivers are near the A-P interface, $1.9-15.8$\,Gbps higher channel capacity is observed compared with the channel capacity corresponding to $1$\,mm and $1.95$\,mm burial depths. This difference is due to the reduced path loss from the LW-A when the transceivers are placed close to the A-P interface and the higher absorption effect of paint and plaster mediums for the other transceiver placements. Even though approximately equal channel capacity is noticed for $1$ and $1.95$\,mm burial depths until $2.3$\,cm LoS distance due to receive power balancing from the multipath, $0.003-1.47$\,Gbps increasing gap is noticed afterward. When comparing predicted channel capacity for air and IoP communication over different burial depths (see Fig. \ref{fig:CC_Air_paint}(b)), channel capacity is reduced between $97$ and $140$\,Gbps, keeping other factors constant. The channel capacity loss in the paint medium is due to increases in both spreading and absorption losses. Note that spreading losses for paint differ from those in air. When waves enter a medium with a higher refractive index, their wavelength is reduced, and so the wave speed is also reduced \cite{salam2017smart}.   

All simulations were performed with the same transmitted power. Therefore, further experiments are needed to assess how effective propagation distance and channel capacity relate to transmitted power.

\section{Conclusion}
\label{sec:Conclusion}

In this article, we introduce a channel model and capacity analysis for nano-networks in the IoP. In this new paradigm shift, nano-network devices are mixed into the paint composition and applied to wall, resulting in a thin film of paint with integrated communication networks. The scale of the devices, which are at the micron scale, will be constructed from components assembled from nanomaterials. In particular, the devices will have nanoantennas built from metamaterials (e.g., graphene) and given its scale will communicate in the THz frequency spectrum. Given the properties of THz waves propagating through the paint medium, as well as between different interfaces, such as plaster and air, we defined a number of critical paths. This includes direct waves, waves reflected from the Air-Paint (A-P) and paint-plaster (P-P) interfaces, as well as lateral waves propagating along both A-P and P-P boundaries. Furthermore, our investigation considers three distinct types (Brilliant Blue, Titanium White, and Oxide Black) of paint in our analysis for path loss and channel capacity. Numerical evaluations indicate that the path loss increases at a moderate rate with both frequency and Line of Sight (LoS) distances between transceivers. Notably, paints with higher refractive indexes result in higher path loss. 

Further, when the nano-devices are buried at similar depths and positioned near the A-P interface, the lateral waves exhibit relatively promising path loss performance with increasing distances between the transmitter and receiver. This is because, even though the path is longer, the waves that propagate to the boundary and propagate along the surface in the air encounter lower loss. Intuitively, increasing the thickness of the paint layer leads to more significant attenuation for the waves propagating within the medium.
Also, the total received signal strength demonstrates an initial steep decline as burial depth increases, before increasing slightly when the transceivers get closer to the P-P interface. The analysis of channel capacity in the different color paints indicates that it reduces exponentially with distance, and applying paints with low refractive index could achieve higher channel capacity. Moreover, one order of magnitude higher channel capacity could be expected by placing transceivers in the proximity of the A-P interface compared to the higher burial depths. Finally, when comparing the IoP communication with the THz propagation on clear air, there is almost two orders of magnitude reduction in the channel capacity. 

The IoP can result in the largest distribution of miniature nano-devices that connect to the cyber-space, where they will seamlessly be embedded into building structures. This can result in novel applications, which are as follows:  
\begin{itemize}
    \item {\bf IoP for green-house gas sensing}: The nano-devices embedded into the paint mixture can use point-to-point THz signaling to sense gases (e.g., greenhouse gas emissions \cite{Wedage_2023}), pathogens, etc. This, in turn, leads to data generated in large geographical areas that will allow us to correlate tracking long-term gas changes for climate change.

    \item \textbf{Posture Recognition}: The sub-THz and THz signals themselves could be used for sensing purposes (i.e., RF-based sensing), where the high bandwidth provides significantly improved accuracy, providing various sensing opportunities such as in-room gaming, posture recognition, head recognition and headset sensing for AR/VR, position sensing, immersive media for metaverse \cite{IF_Akyildiz2023}, etc.

    \item \textbf{Back-scatter type of sensing}: The nanodevices inside the paint can be powered/read using a scanner, such as from an ultrasound transducer. This source can emit ultrasound waves at the nano-devices, and sense the back-scattered waves containing information. This approach eliminates the need for powering individual nodes. Additionally, this can allow the nanodevices to be embedded in the paint for years, only activating when they need to be read. Thus, many high-resolution sensing applications can be developed using an IoP application.
    
    \item {\bf IoP for public health monitoring}: The recent COVID-19 pandemic has led to increased attention for pathogen sensing. The IoP can facilitate the sensing of viral particles, which can result from violent respiratory excretion (e.g., sneezing). Similar to gas sensing, the THz signals can be used to viral particle attachments on the wall. This can lead to large-scale data collection to track the epidemic movement of the virus. 
    
    \item {\bf Wall communication infrastructure}: The IoP can result in a new paint-based IRS, where nano-devices are used to reflect and steer THz signals. At the same time, the nano-networks can allow communication signals to propagate along the wall, acting as a wave guide. This in turn can lay down the fears of the public in terms of wireless signals affecting health as the signals will concentrate and move along the walls rather than propagate in the open space. 
    
\end{itemize}

\section*{Acknowledgments}
This publication came from research conducted with the financial support of Science Foundation Ireland (SFI) and the Department of Agriculture, Food and Marine on behalf of the Government of Ireland (Grant Number [16/RC/3835] - VistaMilk), the support of YL Verkot, Finland, and US National Science Foundation (NSF) ECCS-2030272 and CBET-2316960 grants.

\bibliographystyle{IEEEtran}
\bibliography{References}

\begin{thebibliography}{10}
\providecommand{\url}[1]{#1}
\csname url@samestyle\endcsname
\providecommand{\newblock}{\relax}
\providecommand{\bibinfo}[2]{#2}
\providecommand{\BIBentrySTDinterwordspacing}{\spaceskip=0pt\relax}
\providecommand{\BIBentryALTinterwordstretchfactor}{4}
\providecommand{\BIBentryALTinterwordspacing}{\spaceskip=\fontdimen2\font plus
\BIBentryALTinterwordstretchfactor\fontdimen3\font minus \fontdimen4\font\relax}
\providecommand{\BIBforeignlanguage}[2]{{%
\expandafter\ifx\csname l@#1\endcsname\relax
\typeout{** WARNING: IEEEtran.bst: No hyphenation pattern has been}%
\typeout{** loaded for the language `#1'. Using the pattern for}%
\typeout{** the default language instead.}%
\else
\language=\csname l@#1\endcsname
\fi
#2}}
\providecommand{\BIBdecl}{\relax}
\BIBdecl

\bibitem{Nano_network_2021}
F.~Lemic, S.~Abadal, W.~Tavernier, P.~Stroobant, D.~Colle, E.~Alarcón, J.~Marquez-Barja, and J.~Famaey, ``Survey on terahertz nanocommunication and networking: A top-down perspective,'' \emph{IEEE Journal on Selected Areas in Communications}, vol.~39, no.~6, pp. 1506--1543, 2021.

\bibitem{llatser2013graphene}
I.~Llatser, S.~Abadal, A.~M. Sugranes, A.~Cabellos-Aparicio, and E.~Alarc{\'o}n, ``Graphene-enabled wireless networks-on-chip,'' in \emph{2013 First International Black Sea Conference on Communications and Networking (BlackSeaCom)}.\hskip 1em plus 0.5em minus 0.4em\relax IEEE, 2013, pp. 69--73.

\bibitem{elayan2017terahertz}
H.~Elayan, R.~M. Shubair, J.~M. Jornet, and P.~Johari, ``Terahertz channel model and link budget analysis for intrabody nanoscale communication,'' \emph{IEEE transactions on nanobioscience}, vol.~16, no.~6, pp. 491--503, 2017.

\bibitem{vizziello2023intra}
A.~Vizziello, M.~Magarini, P.~Savazzi, and L.~Galluccio, ``Intra-body communications for nervous system applications: Current technologies and future directions,'' \emph{Computer Networks}, vol. 227, p. 109718, 2023.

\bibitem{jornet2023nanonetworking}
J.~M. Jornet and A.~Sangwan, ``Nanonetworking in the terahertz band and beyond,'' \emph{IEEE Nanotechnology Magazine}, 2023.

\bibitem{Wedage_2023}
L.~T. Wedage, B.~Butler, S.~Balasubramaniam, Y.~Koucheryavy, J.~M. Jornet, and M.~C. Vuran, ``Climate change sensing through terahertz communication infrastructure: A disruptive application of 6{G} networks,'' \emph{IEEE Network}, pp. 1--1, 2023.

\bibitem{THz_detectors}
F.~Sizov and A.~Rogalski, ``Thz detectors,'' \emph{Progress in quantum electronics}, vol.~34, no.~5, pp. 278--347, 2010.

\bibitem{kianoush2019passive}
S.~Kianoush, S.~Savazzi, and V.~Rampa, ``Passive detection and discrimination of body movements in the sub-thz band: A case study,'' in \emph{ICASSP 2019-2019 IEEE International Conference on Acoustics, Speech and Signal Processing (ICASSP)}.\hskip 1em plus 0.5em minus 0.4em\relax IEEE, 2019, pp. 1597--1601.

\bibitem{dai2009remote}
J.~Dai, X.~Lu, J.~Liu, I.~Ho, N.~Karpowicz, and X.~Zhang, ``Remote thz wave sensing in ambient atmosphere,'' \emph{Science}, vol.~2, pp. 131--143, 2009.

\bibitem{sensing2019room}
A.~Sample, C.~Yang, and Y.~Zhang, ``Room-scale interactive and context-aware sensing,'' Jul.~16 2019, uS Patent 10,353,526.

\bibitem{pappa2011oldest}
S.~Pappa, ``Oldest human paint-making studio discovered in cave,'' \emph{Live Science}, 2011.

\bibitem{sample2011stone}
I.~Sample, ``Stone age painting kits found in cave [online]. uk: Guardian newspaper,'' 2011.

\bibitem{cencillo2023ultralight}
P.~Cencillo-Abad, D.~Franklin, P.~Mastranzo-Ortega, J.~Sanchez-Mondragon, and D.~Chanda, ``Ultralight plasmonic structural color paint,'' \emph{Science Advances}, vol.~9, no.~10, p. eadf7207, 2023.

\bibitem{gomez2023optimizing}
J.~T. G{\'o}mez, J.~Simonjan, J.~M. Jornet, and F.~Dressler, ``Optimizing terahertz communication between nanosensors in the human cardiovascular system and external gateways,'' \emph{IEEE Communications Letters}, 2023.

\bibitem{reddy2023photothermal}
I.~V.~K. Reddy, S.~Elmaadawy, E.~P. Furlani, and J.~M. Jornet, ``Photothermal effects of terahertz-band and optical electromagnetic radiation on human tissues,'' \emph{Scientific Reports}, vol.~13, no.~1, p. 14643, 2023.

\bibitem{timoneda2018millimeter}
X.~Timoneda, S.~Abadal, A.~Cabellos-Aparicio, D.~Manessis, J.~Zhou, A.~Franques, J.~Torrellas, and E.~Alarc{\'o}n, ``Millimeter-wave propagation within a computer chip package,'' in \emph{2018 IEEE International Symposium on Circuits and Systems (ISCAS)}.\hskip 1em plus 0.5em minus 0.4em\relax IEEE, 2018, pp. 1--5.

\bibitem{abadal2019opportunistic}
S.~Abadal, A.~Marruedo, A.~Franques, H.~Taghvaee, A.~Cabellos-Aparicio, J.~Zhou, J.~Torrellas, and E.~Alarc{\'o}n, ``Opportunistic beamforming in wireless network-on-chip,'' in \emph{2019 IEEE International Symposium on Circuits and Systems (ISCAS)}.\hskip 1em plus 0.5em minus 0.4em\relax IEEE, 2019, pp. 1--5.

\bibitem{abadal2019wave}
S.~Abadal, C.~Han, and J.~M. Jornet, ``Wave propagation and channel modeling in chip-scale wireless communications: A survey from millimeter-wave to terahertz and optics,'' \emph{IEEE access}, vol.~8, pp. 278--293, 2019.

\bibitem{vuran2018internet}
M.~C. Vuran, A.~Salam, R.~Wong, and S.~Irmak, ``Internet of underground things in precision agriculture: Architecture and technology aspects,'' \emph{Ad Hoc Networks}, vol.~81, pp. 160--173, 2018.

\bibitem{dworak2017terahertz}
V.~Dworak, B.~Mahns, J.~Selbeck, R.~Gebbers, and C.~Weltzien, ``Terahertz spectroscopy for proximal soil sensing: An approach to particle size analysis,'' \emph{Sensors}, vol.~17, no.~10, p. 2387, 2017.

\bibitem{kaushal2016underwater}
H.~Kaushal and G.~Kaddoum, ``Underwater optical wireless communication,'' \emph{IEEE access}, vol.~4, pp. 1518--1547, 2016.

\bibitem{Wedage_comparative_2023}
L.~T. Wedage, B.~Butler, S.~Balasubramaniam, Y.~Koucheryavy, and M.~C. Vuran, ``Comparative analysis of terahertz propagation under dust storm conditions on mars and earth,'' \emph{IEEE Journal of Selected Topics in Signal Processing}, pp. 1--16, 2023.

\bibitem{hossain2019stochastic}
Z.~Hossain, C.~N. Mollica, J.~F. Federici, and J.~M. Jornet, ``Stochastic interference modeling and experimental validation for pulse-based terahertz communication,'' \emph{IEEE Transactions on Wireless Communications}, vol.~18, no.~8, pp. 4103--4115, 2019.

\bibitem{monnai2023terahertz}
Y.~Monnai, X.~Lu, and K.~Sengupta, ``Terahertz beam steering: from fundamentals to applications,'' \emph{Journal of Infrared, Millimeter, and Terahertz Waves}, vol.~44, no. 3-4, pp. 169--211, 2023.

\bibitem{wang2008towards}
Z.~L. Wang, ``Towards self-powered nanosystems: from nanogenerators to nanopiezotronics,'' \emph{Advanced Functional Materials}, vol.~18, no.~22, pp. 3553--3567, 2008.

\bibitem{xu2010piezoelectric}
S.~Xu, B.~J. Hansen, and Z.~L. Wang, ``Piezoelectric-nanowire-enabled power source for driving wireless microelectronics,'' \emph{Nature communications}, vol.~1, no.~1, p.~93, 2010.

\bibitem{seo2016wireless}
D.~Seo, R.~M. Neely, K.~Shen, U.~Singhal, E.~Alon, J.~M. Rabaey, J.~M. Carmena, and M.~M. Maharbiz, ``Wireless recording in the peripheral nervous system with ultrasonic neural dust,'' \emph{Neuron}, vol.~91, no.~3, pp. 529--539, 2016.

\bibitem{khan2020high}
M.~A.~K. Khan, M.~I. Ullah, R.~Kabir, and M.~A. Alim, ``High-performance graphene patch antenna with superstrate cover for terahertz band application,'' \emph{Plasmonics}, vol.~15, pp. 1719--1727, 2020.

\bibitem{salam2017smart}
A.~Salam and M.~C. Vuran, ``Smart underground antenna arrays: A soil moisture adaptive beamforming approach,'' in \emph{IEEE INFOCOM 2017-IEEE conference on computer communications}.\hskip 1em plus 0.5em minus 0.4em\relax IEEE, 2017, pp. 1--9.

\bibitem{jornet2011channel}
J.~M. Jornet and I.~F. Akyildiz, ``Channel modeling and capacity analysis for electromagnetic wireless nanonetworks in the terahertz band,'' \emph{IEEE Transactions on Wireless Communications}, vol.~10, no.~10, pp. 3211--3221, 2011.

\bibitem{Akyildiz2012terahertz}
M.~A. Akka{\c{s}}, I.~F. Akyildiz, and R.~Sokullu, ``Terahertz channel modeling of underground sensor networks in oil reservoirs,'' in \emph{2012 IEEE Global Communications Conference (GLOBECOM)}.\hskip 1em plus 0.5em minus 0.4em\relax IEEE, 2012, pp. 543--548.

\bibitem{refractive2008}
C.~Jansen, R.~Piesiewicz, D.~Mittleman, T.~Kurner, and M.~Koch, ``The impact of reflections from stratified building materials on the wave propagation in future indoor terahertz communication systems,'' \emph{IEEE Transactions on Antennas and Propagation}, vol.~56, no.~5, pp. 1413--1419, 2008.

\bibitem{salam2020statistical}
A.~Salam, M.~C. Vuran, and S.~Irmak, ``A statistical impulse response model based on empirical characterization of wireless underground channels,'' \emph{IEEE Transactions on Wireless Communications}, vol.~19, no.~9, pp. 5966--5981, 2020.

\bibitem{han2014multi}
C.~Han, A.~O. Bicen, and I.~F. Akyildiz, ``Multi-ray channel modeling and wideband characterization for wireless communications in the terahertz band,'' \emph{IEEE Transactions on Wireless Communications}, vol.~14, no.~5, pp. 2402--2412, 2014.

\bibitem{tsujimura2017causal}
K.~Tsujimura, K.~Umebayashi, J.~Kokkoniemi, J.~Lehtom{\"a}ki, and Y.~Suzuki, ``A causal channel model for the terahertz band,'' \emph{IEEE Transactions on Terahertz Science and Technology}, vol.~8, no.~1, pp. 52--62, 2017.

\bibitem{piesiewicz2007scattering}
R.~Piesiewicz, C.~Jansen, D.~Mittleman, T.~Kleine-Ostmann, M.~Koch, and T.~Kurner, ``Scattering analysis for the modeling of thz communication systems,'' \emph{IEEE Transactions on Antennas and Propagation}, vol.~55, no.~11, pp. 3002--3009, 2007.

\bibitem{Clough76}
J.~W. Clough, ``Electromagnetic lateral waves observed by earth sounding radars,'' \emph{Geophysics}, vol.~41, no.~6A, pp. 1126--1132, Dec 1976.

\bibitem{Wu82}
T.~T. Wu and R.~W.~P. King, ``Lateral waves: A new formula and interference patterns,'' \emph{Radio Science}, vol.~17, no.~3, pp. 521--531, May-June 1982.

\bibitem{Ginzel08}
E.~Ginzel, F.~Honarvar, and A.~Yaghootian, ``A study of time-of-flight diffraction technique using photoelastic visualisations,'' in \emph{Int. Conf. Technical Inspection and NDT}, Oct 2008.

\bibitem{Dence69}
D.~Dence and T.~Tamir, ``Radio loss of lateral waves in forest environments,'' \emph{Radio Science}, vol.~4, no.~4, pp. 307--318, April 1969.

\bibitem{Hitran2016}
I.~Gordon \emph{et~al.}, ``\BIBforeignlanguage{en}{The {HITRAN2016} {Molecular Spectroscopic Database}},'' \emph{\BIBforeignlanguage{en}{Journal of Quantitative Spectroscopy and Radiative Transfer}}, vol. 203, pp. 3--69, Dec. 2017.

\bibitem{Wedage_pathloss_2022}
L.~T. Wedage, B.~Butler, S.~Balasubramaniam, M.~C. Vuran, and Y.~Koucheryavy, ``Path loss analysis of terahertz communication in mars' atmospheric conditions,'' in \emph{2022 IEEE International Conference on Communications Workshops (ICC Workshops)}, 2022, pp. 1225--1230.

\bibitem{ep_value}
Z.~Diao, Q.~Jing, and W.~Zhong, ``Comparison of the influence of martian and earth's atmospheric environments on terahertz band electromagnetic waves,'' \emph{International Journal of Communication Systems}, vol.~34, no.~12, p. e4894, 2021.

\bibitem{dong2011channel}
X.~Dong and M.~C. Vuran, ``A channel model for wireless underground sensor networks using lateral waves,'' in \emph{2011 IEEE Global Telecommunications Conference-GLOBECOM 2011}.\hskip 1em plus 0.5em minus 0.4em\relax IEEE, 2011, pp. 1--6.

\bibitem{goldsmith2005wireless}
A.~Goldsmith, \emph{Wireless communications}.\hskip 1em plus 0.5em minus 0.4em\relax Cambridge university press, 2005.

\bibitem{jornet2010capacity}
J.~M. Jornet and I.~F. Akyildiz, ``Channel capacity of electromagnetic nanonetworks in the terahertz band,'' in \emph{2010 IEEE international conference on communications}.\hskip 1em plus 0.5em minus 0.4em\relax IEEE, 2010, pp. 1--6.

\bibitem{da2009carbon}
M.~R. Da~Costa, O.~Kibis, and M.~Portnoi, ``Carbon nanotubes as a basis for terahertz emitters and detectors,'' \emph{Microelectronics Journal}, vol.~40, no. 4-5, pp. 776--778, 2009.

\bibitem{King92Book}
R.~King, M.~Owens, and T.~T. Wu, \emph{Lateral Electromagnetic Waves: Theory and Applications to Communications, Geophysical Exploration, and Remote Sensing}.\hskip 1em plus 0.5em minus 0.4em\relax Springer-Verlag, 2092.

\bibitem{Wabia92}
M.~Wabia, ``Lateral waves in anisotropic optical waveguides,'' \emph{Acta Physica Polonica A}, vol.~81, no. 4-5, pp. 503--516, 1992.

\bibitem{Beer-Lam_1995}
R.~M. Goody and Y.~L. Yung, \emph{Atmospheric radiation: theoretical basis}.\hskip 1em plus 0.5em minus 0.4em\relax Oxford university press, 1995.

\bibitem{abraham2010non}
E.~Abraham, A.~Younus, J.-C. Delagnes, and P.~Mounaix, ``Non-invasive investigation of art paintings by terahertz imaging,'' \emph{Applied Physics A}, vol. 100, pp. 585--590, 2010.

\bibitem{otoshi1999measurements}
T.~Otoshi, R.~Cirillo~Jr, and J.~Sosnowski, ``Measurements of complex dielectric constants of paints and primers for dsn antennas: Part i,'' \emph{TMO Progress Rep}, pp. 42--138, 1999.

\bibitem{IF_Akyildiz2023}
I.~F. Akyildiz, H.~Guo, R.~Dai, and W.~Gerstacker, ``Mulsemedia communication research challenges for metaverse in 6{G} wireless systems,'' \emph{ITU Journal on Future and Evolving Technologies}, 2023.

\end{thebibliography}


\begin{IEEEbiography}[{\includegraphics[width=1in,height=1.25in,clip,keepaspectratio]{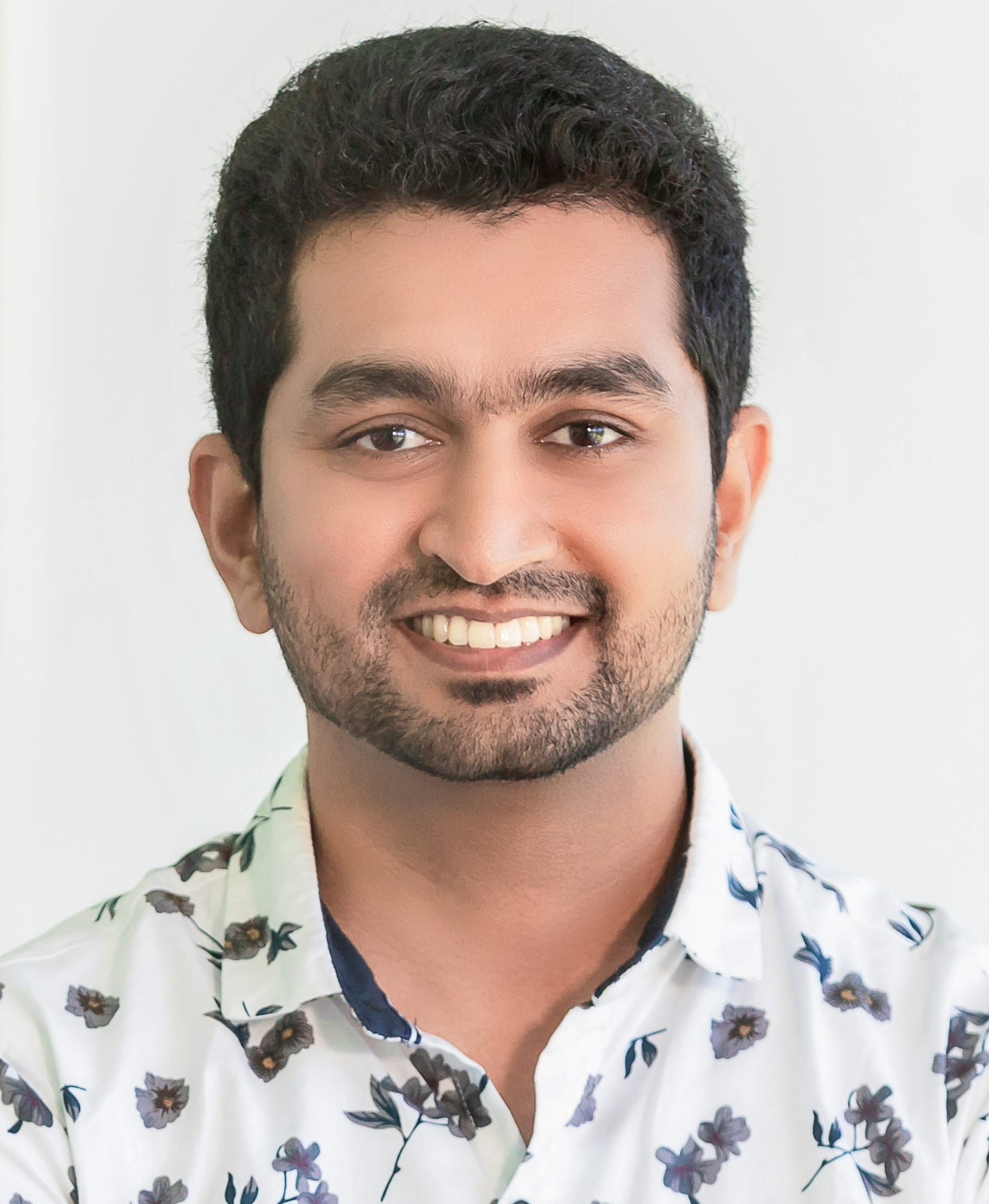}}] {LASANTHA THAKSHILA WEDAGE} [S'22] (thakshila.wedage@waltoninstitute.ie)
received his B.S. degree in Mathematics from University of Ruhuna, Sri Lanka, in 2016. He is currently pursuing a PhD degree with the Department of Computing and Mathematics at South East Technological University, Ireland. His current research interests include Mathematical modeling and statistics, 5G/6G Wireless communication, and sensing.
\end{IEEEbiography}

\begin{IEEEbiography}[{\includegraphics[width=1in,height=1.25in,clip,keepaspectratio]{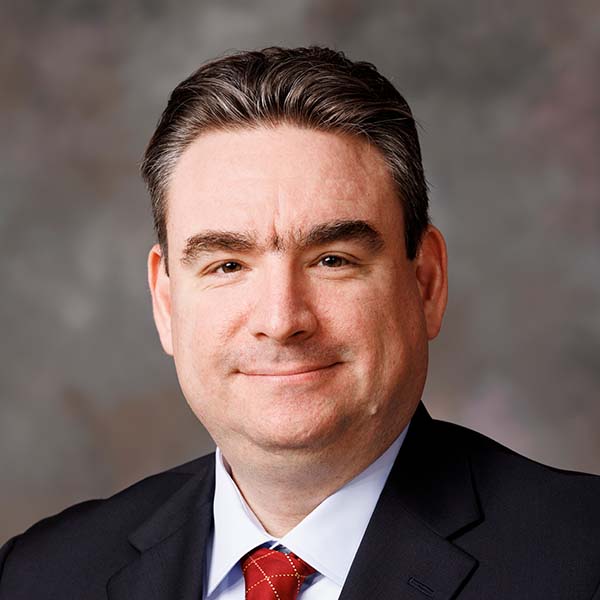}}] {MEHMET CAN VURAN}[M'07] (mcv@unl.edu) received his B.Sc. degree in Electrical and Electronics Engineering from Bilkent University, Ankara, Turkey in 2002. He received his M.S. and Ph.D. degrees in Electrical and Computer Engineering from Georgia Institute of Technology, Atlanta, GA. in 2004 and 2007, respectively. He is currently the Dale M. Jensen Professor of Computing with the University of Nebraska-Lincoln. He is a Daugherty Water for Food Institute Fellow and a National Strategic Research Institute Fellow. He was awarded an NSF CAREER Award for the project “Bringing Wireless Sensor Networks Underground.” His research interests include wireless underground, mmWave, and THz communications in challenging environments, agricultural Internet of Things, dynamic spectrum access, wearable embedded systems, connected autonomous systems, and cyber-physical networking.
\end{IEEEbiography}

\begin{IEEEbiography}[{\includegraphics[width=1in,height=1.25in,clip,keepaspectratio]{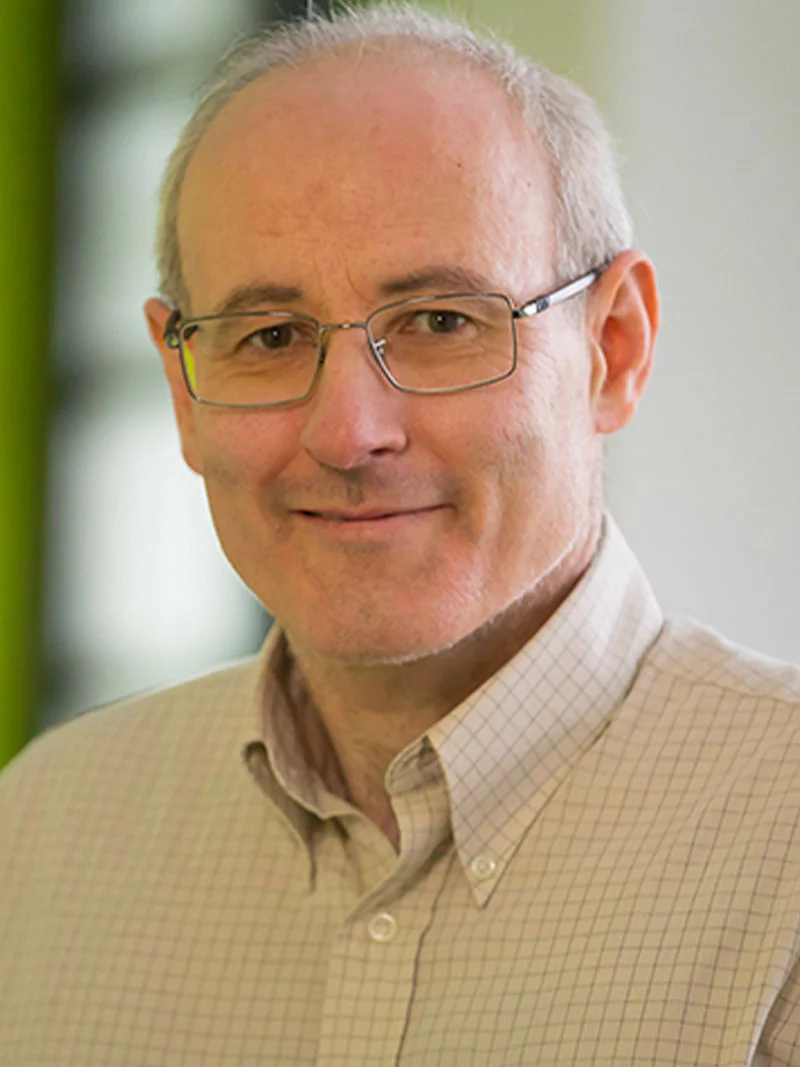}}]{BERNARD BUTLER}[SM'22] (bbutler@ieee.org) received a PhD from Waterford Institute of Technology, Ireland. He was a Senior Research Scientist in the U.K.’s National Physical Laboratory, focusing on mathematics of measurement and sensing. He is a lecturer in SETU and a CONNECT Funded Investigator and VistaMilk Academic Collaborator in SETU. His research interests include the management of distributed computing and sensing systems, applied to future networking and agriculture.
\end{IEEEbiography}

\begin{IEEEbiography}[{\includegraphics[width=1in,height=1.25in,clip,keepaspectratio]{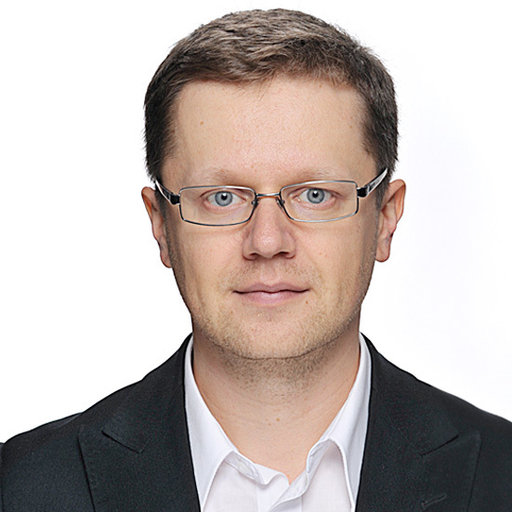}}]{YEVGENI KOUCHERYAVY}[SM'08] (yevgeni.koucheryavy@yl-verkot.com) received the Ph.D. degree from the Tampere University of Technology, Finland, in 2004. He is currently a Full Professor with the Unit of Electrical Engineering, Tampere University, Finland. He has authored numerous publications in the field of advanced wired and wireless networking and communications. His current research interests include various aspects in heterogeneous wireless communication networks and systems, the Internet of Things and its standardization, and nanocommunications.
\end{IEEEbiography}

\begin{IEEEbiography}[{\includegraphics[width=1in,height=1.25in,clip,keepaspectratio]{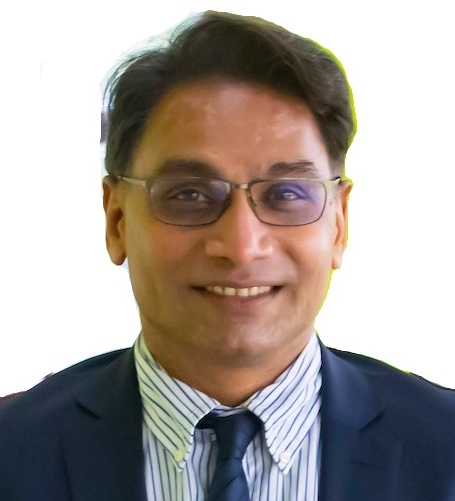}}]{SASITHARAN BALASUBRAMANIAM} [SM'14] (sasi@unl.edu) received the Bachelor’s degree in engineering and Ph.D. degree from University of Queensland, Brisbane, Australia, in 1998 and 2005, respectively, and the Masters of engineering science from Queensland University of Technology, Brisbane, in 1999. He was a recipient of Science Foundation Ireland Starter Investigator Research Grant. He was also a recipient of the Academy of Finland Research Fellow with Tampere University, Finland. He was the Director of Research at the Walton Institute, South East Technological University, Ireland. He is currently an Associate Professor with the School of Computing, University of Nebraska-Lincoln, Lincoln, NE, USA. His research interests include molecular/nano communications, Internet of Bio-Nano Things, and 5G/6G networks. He is currently the Editor-in-Chief of IEEE Transactions on Molecular, Biological and Multi-Scale Communications and an Associate Editor for IEEE Transactions on Mobile Computing and IEEE Transactions on Nanobioscience. He was an IEEE Distinguished Lecturer for the IEEE Nanotechnology Council in 2018.
\end{IEEEbiography}

\end{document}